\documentclass[twocolumn,english,aps,english,showpacs,preprintnumbers,groupedaddress,amsmath,amssymb,pra]{revtex4}
\usepackage[T1]{fontenc}
\usepackage[latin9]{inputenc}
\usepackage{amsmath}
\usepackage{amssymb}
\usepackage{graphicx}
\usepackage{esint}

\makeatletter
\@ifundefined{textcolor}{}
{%
 \definecolor{BLACK}{gray}{0}
 \definecolor{WHITE}{gray}{1}
 \definecolor{RED}{rgb}{1,0,0}
 \definecolor{GREEN}{rgb}{0,1,0}
 \definecolor{BLUE}{rgb}{0,0,1}
 \definecolor{CYAN}{cmyk}{1,0,0,0}
 \definecolor{MAGENTA}{cmyk}{0,1,0,0}
 \definecolor{YELLOW}{cmyk}{0,0,1,0}
}

\@ifundefined{textcolor}{}{%
 \definecolor{BLACK}{gray}{0}
 \definecolor{WHITE}{gray}{1}
 \definecolor{RED}{rgb}{1,0,0}
 \definecolor{GREEN}{rgb}{0,1,0}
 \definecolor{BLUE}{rgb}{0,0,1}
 \definecolor{CYAN}{cmyk}{1,0,0,0}
 \definecolor{MAGENTA}{cmyk}{0,1,0,0}
 \definecolor{YELLOW}{cmyk}{0,0,1,0}
}

\makeatother

\usepackage{babel}
\begin{document}

\title{Simple asymptotic forms for Sommerfeld and Brillouin precursors}

\author{Bruno Macke}

\author{Bernard S\'{e}gard}

\email{bernard.segard@univ-lille-1.fr}

\affiliation{Laboratoire de Physique des Lasers, Atomes et Mol\'{e}cules , CNRS et
Universit\'{e} Lille 1, 59655 Villeneuve d'Ascq, France}

\date{\today}
\begin{abstract}
This article mainly deals with the propagation of step-modulated light
pulses in a dense Lorentz-medium at distances such that the medium
is opaque in a broad spectral region including the carrier frequency.
The transmitted field is then reduced to the celebrated precursors
of Sommerfeld and Brillouin, far apart from each other. We obtain
simple analytical expressions of the first (Sommerfeld) precursor
whose shape only depends on the order of the initial discontinuity
of the incident field and whose amplitude rapidly decreases with this
order (rise-time effects). We show that, in a strictly speaking asymptotic
limit, the second (Brillouin) precursor is entirely determined by
the frequency-dependence of the medium attenuation and has a Gaussian
or Gaussian-derivative shape. We point out that this result applies
to the precursor directly observed in a Debye medium at decimetric
wavelengths. When attenuation and group-delay dispersion both contribute
to its formation, we establish a more general expression of the Brillouin
precursor, containing the previous one (dominant-attenuation limit)
and that obtained by Brillouin (dominant-dispersion limit) as particular
cases. We finally study the propagation of square or Gaussian pulses
and we determine the pulse parameters optimizing the Brillouin precursor.
Obtained by standard Laplace-Fourier procedures, our results are explicit
and contrast by their simplicity from those derived by the uniform
saddle point methods, from which it is very difficult to retrieve
our asymptotic forms. 
\end{abstract}


\maketitle

\section{INTRODUCTION\label{sec:INTRODUCTION}}

More than one century ago, in a short communication \cite{som07}
made at the $79^{\mathrm{th}}$ congress of the German physicists,
Sommerfeld examined the apparent inconsistency between the theory
of special relativity and the possibility of superluminal group velocity
predicted by the classical wave theory. Considering an incident wave
switched on at time $t=-T$ and off at time $t=T$ (square-wave modulation),
he mathematically demonstrated that, regardless of the value of the
group velocity at the frequency of the optical carrier, no signal
can be transmitted by any linear dispersive-attenuative medium before
the instant $t=-T+z/c$, where $z$ is the propagation distance and
$c$ the velocity of light in vacuum. In the discussion following
the Sommerfeld's communication, Voigt proposed a simple physical interpretation
of this result. He remarked that the front of the wave encounters
a medium that, due to its inertia, seems optically empty and, thus,
that the propagation of the very first beginning of the signal will
proceed undisturbed with the velocity of light in vacuum. In other
words, local causality implies relativistic causality. The analysis
of what happens after the arrival of the wavefront was subsequently
conducted by Sommerfeld and Brillouin in the case of a step-wave modulation
(field switched on at time $t=0$), the medium being modeled as an
ensemble of damped harmonic oscillators with the same resonance frequency
$\omega_{0}$ and the same damping rate $\gamma$ (Lorentz medium)
\cite{som14,bri14,bri32,bri60}. They found that, in suitable conditions,
the transmitted signal consists in two successive transients (that
they named ``forerunners\textquotedblright{}) preceding the establishment
of the steady-state field at the frequency $\omega_{c}$ of the optical
carrier (the ``main field\textquotedblright{}). The first and second
forerunners, now called the Sommerfeld and Brillouin precursors, were
associated with the frequencies respectively high and low compared
to the resonance frequency $\omega_{0}$ of the medium. These results
were obtained by means of a spectral approach involving the newly
developed saddle-point method \cite{bri14} but also classical complex
analysis \cite{som14} and stationary phase method \cite{bri32}.
Following these pioneering works, precursors became a canonical problem
in electromagnetism and optics \cite{stra41,ja75}. Results completing,
improving and even correcting those of Sommerfeld and Brillouin were
obtained by means of uniform asymptotic methods \cite{han69,ou75,va88,ou89}.
The problem was also studied by a purely temporal approach \cite{ka98}.
At the present time, the theoretical study of precursors continues
to raise a considerable interest. An abundant bibliography can be
found in the recent Oughstun's book \cite{ou09}. Complementary studies
on the effects of a finite turn-on time of the incident field on the
precursors are reported in \cite{cia02,cia03,bm11,cia11}.

From an experimental point of view, the observation of Sommerfeld
and Brillouin precursors \emph{in the optical range} raises serious
difficulties. Indeed the excitation of the Sommerfeld and Brillouin
precursors requires the corresponding frequencies (respectively high
and low compared to $\omega_{0}$) be present at a significant level
in the spectrum of the incident pulse. An experiment intended to observe
the Brillouin precursor in water is reported in \cite{choi04}. Using
pulses at a wavelength of 700 nm with a bandwidth of 60 nm, the authors
observed pulse breakup in a linear regime as well as a sub-exponential
attenuation with distance of the new peak. They attributed these features
to the formation of a Brillouin precursor. This interpretation has
been soundly disputed, in particular because the pulse bandwidth was
in fact not broad enough to perform the excitation of precursors \cite{alf05}.
Alternative explanations of the observations have been proposed \cite{alf05,ro04}
and more recent studies \cite{lu09,na09,spr11} have confirmed that
a sub-exponential decay of the transmitted energy does not prove the
formation of precursors.

While well distinguishable Sommerfeld and Brillouin precursors are
expected when the medium is opaque in a broad spectral region, coherent
transients of another kind are obtained in the opposite case where
the width of the opacity region is very small compared to the resonance
frequency $\omega_{0}$. They have been naturally named resonant precursors
\cite{va86} but also Sommerfeld-Brillouin precursors \cite{aa91}.
Indeed they may be seen as resulting from the coalescence of the Sommerfeld
and Brillouin precursors, originating a well-marked beat when the
optical thickness of the medium is large enough \cite{bs87}. The
conditions required to achieve experimental evidence of these precursors
are relatively easy to meet. They have been actually observed in various
systems, in particular in a molecular gas \cite{bs87}, in a solid-state
sample with a narrow exciton line \cite{aa91} and in clouds of cold
atoms \cite{jeon06,wei09}.

In the present paper we come back to the study of Sommerfeld and Brillouin
precursors in a dense Lorentz medium, considering the limit where
the medium is opaque in a spectral region of width large compared
to the resonance frequency. We remark that\emph{ these conditions
are met for the parameters considered by Brillouin} \cite{re1} and
often referred to in the literature. We then succeed in obtaining
\emph{simple and explicit analytical expressions} of both precursors.
When it is necessary, we determine the range of validity of these
analytical solutions by comparing them to exact numerical solutions
obtained by fast Fourier transform (FFT). The arrangement of our paper
is as follows. In Section \ref{sec:GENERAL-ANALYSIS}, we outline
the problem under consideration and give some general results, useful
for the following. Section \ref{sec:SOMMERFELD-PRECURSOR} is devoted
to the study of the Sommerfeld precursor. We establish the corresponding
expression of the impulse response of the medium and apply it to obtain
a general expression of the precursors obtained with causal incident
fields. We examine in detail the particular cases where the incident
field is discontinuous at the initial time or has the canonical form
considered by Brillouin with eventually a finite rise time. We show
in Section \ref{sec:BrillouinStrictLimit} that, in a strictly speaking
asymptotic limit, the impulse response associated with the Brillouin
precursor is Gaussian and that the Brillouin precursor has itself
a Gaussian or Gaussian-derivative shape. The precursor obtained in
a Debye medium is incidentally examined. A more general expression
of the Brillouin precursor in the Lorentz medium is established in
Section \ref{sec:EXTENDED-EXPRESSION-BP}, containing the previous
one and that obtained by Brillouin as particular cases. The propagation
in both media of pulses with a square or Gaussian envelope is finally
examined in Section \ref{sec:SQUARE-GAUSS-PULSES} and we determine
the pulse parameters optimizing the Brillouin precursor. We conclude
in Section \ref{sec:SUMMARY} by summarizing and discussing our main
results.

\section{GENERAL ANALYSIS\label{sec:GENERAL-ANALYSIS}}

We consider a one-dimensional optical wave propagating in a Lorentz
medium in the $z$-direction, with an electric field linearly polarized
in the $x$-direction ($x,y,z$ : Cartesian coordinates). We denote
$e(0,t)$ the algebraic amplitude of the field at time $t$ for $z=0$
(inside the medium) and $e(z,t)$ its value after a propagation distance
$z$ through the medium. The incident field $e(0,t)$ being given,
the problem is to determine the transmitted field $e(z,t)$. We take
for $e(0,t)$ the general form 
\begin{equation}
e(0,t)=u(t)\cos(\omega_{c}t-\varphi),\label{eq:zero}
\end{equation}
 including as particular cases the different forms considered in the
literature. $\omega_{c}$ is the frequency of the optical carrier,
$\varphi$ is the phase (eventually time-depending) and $u(t)\geq0$
is the amplitude modulation or field envelope. On the other hand the
medium is fully characterized in the frequency domain by its transfer
function $H(z,\omega)$ relating the Fourier transform $E(z,\omega)$
of $e(z,t)$ to that $E(0,\omega)$ of $e(0,t)$ \cite{pap87}. 
\begin{equation}
E(z,\omega)=H(z,\omega)E(0,\omega).\label{eq:un}
\end{equation}
 In all the following, we take for $t$ a retarded time equal to the
real time minus the luminal propagation time $z/c$ (retarded-time
picture). $H(z,\omega)$ then reads 
\begin{equation}
H(z,\omega)=\exp\left\{ -i\frac{\omega z}{c}\left[\widetilde{n}(\omega)-1\right]\right\} .\label{eq:deux}
\end{equation}
 Here $\widetilde{n}(\omega)$ is the complex refractive index of
the medium at the frequency $\omega$, that is for the Lorentz medium
\begin{equation}
\widetilde{n}(\omega)=\left(1-\frac{\omega_{p}^{2}}{\omega^{2}-\omega_{0}^{2}-2i\gamma\omega}\right)^{1/2},\label{eq:trois}
\end{equation}
 where $\omega_{0}$ is the resonance frequency, $\gamma$ is the
damping or relaxation rate and $\omega_{p}$ is the so-called plasma
frequency whose square is proportional to the number density of absorbers.
$\Re\left[\widetilde{n}(\omega)\right]$ is the usual (real) refractive
index $n(\omega)$ and the absorption coefficient $\alpha(\omega)$
\emph{for the amplitude} is given by the relation $\alpha(\omega)=-(\omega/c)\Im\left[\widetilde{n}(\omega)\right]$.

In the time domain, the medium will be characterized by its impulse
response $h(z,t)$, inverse Fourier transform of $H(z,\omega)$, and
the transmitted signal $e(z,t)$ is given by the convolution product
\cite{pap87} 
\begin{equation}
e(z,t)=h(z,t)\otimes e(0,t).\label{eq:quatre}
\end{equation}
 Some general properties of $h(z,t)$ and $e(z,t)$ can be deduced
from Eqs.(\ref{eq:deux}-\ref{eq:quatre}). First $h(z,t)$ fulfills
the condition of relativistic causality, namely $h(z,t)=0$ for $t<0$
\cite{re2}. Its area reads as $\intop_{-\infty}^{+\infty}h(z,t)dt=H(z,0)=1$.
It keeps thus constant and normalized to unity regardless of the propagation
distance $z$. Consequently $E(z,0)=E(0,0)$ , that is 
\begin{equation}
\intop_{-\infty}^{+\infty}e(z,t)dt=\intop_{-\infty}^{+\infty}e(0,t)dt.\label{eq:cinq}
\end{equation}
 The area of the optical field (to distinguish from that of its envelope)
is conserved during the propagation. Finally the fact that $H(z,\infty)=1$
entails that $h(z,t)$ will start by a Dirac delta-function $\delta(t)$.
This implies that the propagation of the very first beginning of any
incident signal $e(0,t)$ will always proceed undisturbed at the velocity
$c$, in agreement with the Voigt's remark on the Sommerfeld's communication
\cite{som07}. The previous results are valid whatever the values
of the parameters may be.

Examine now in what conditions the medium is opaque in a broad spectral
region. To be definite, we will consider that the medium is opaque
at the frequency $\omega$ when its optical thickness $\alpha(\omega)z$
exceeds $20$, the amplitude transmission $\left|H(z,\omega)\right|=\exp\left[-\alpha(\omega)z\right]$
being then about $2\times10^{-9}$. Following Sommerfeld \cite{som14},
we characterize the propagation distance by the parameter $\xi=\omega_{p}^{2}z/2c$
, homogeneous to a frequency. For large propagation distances $\gamma\xi/(10\omega_{0}^{2})\gg1$
and it is easily derived from Eq. \ref{eq:trois} that the medium
will then be opaque in the broad spectral region $\omega_{-}\leq\omega\leq\omega_{+}$
with $\omega_{+}/\omega_{0}\approx\sqrt{\gamma\xi/(10\omega_{0}^{2})}$
and $\omega_{-}/\omega_{0}\approx\left(1+\omega_{p}^{2}/\omega_{0}^{2}\right)^{1/4}\sqrt{10\omega_{0}^{2}/(\gamma\xi)}$.
The inequality $\gamma\xi/(10\omega_{0}^{2})\gg1$ is over-satisfied
for the parameters values considered by Brillouin \cite{re1}, namely
$\omega_{0}=4\times10^{16}\:\mathrm{s}^{-1}$, $\omega_{p}^{2}=1.24\:\omega_{0}^{2}$,
$\gamma^{2}=\omega_{0}^{2}/200$ and $z=10^{-2}\:\mathrm{m}$. We
then get $\xi=3.0324\times10^{21}\mathrm{s^{-1}}$ and $\gamma\xi/(10\omega_{0}^{2})\approx5.87\times10^{3}$.
Not to reduce our study to a particular system or region of the spectrum,
all the frequencies (the times) will be referred in the following
to their natural unit $\omega_{0}$ ($1/\omega_{0}$). 
\begin{figure}[h]
\begin{centering}
\includegraphics[width=80mm]{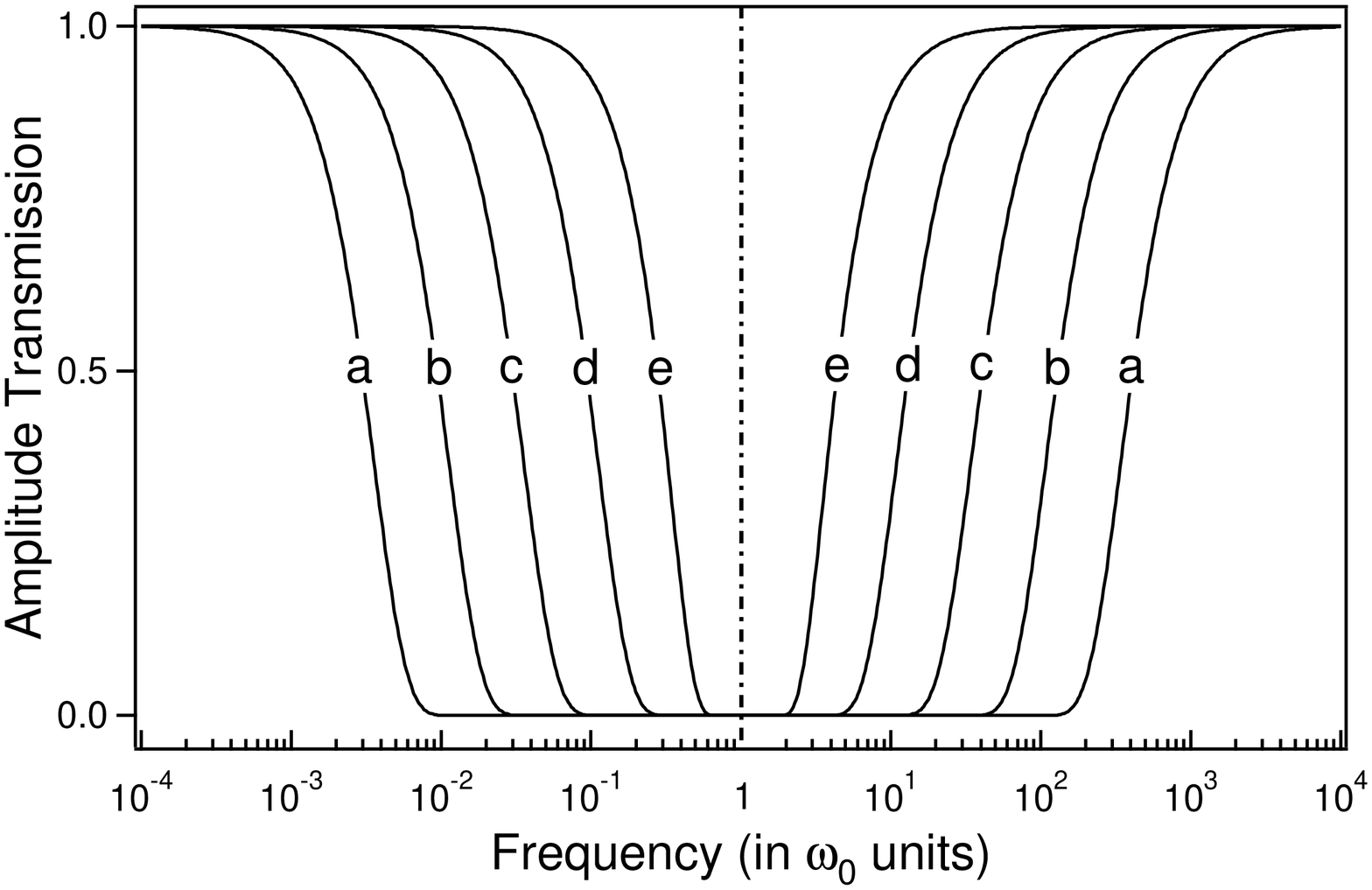} 
\par\end{centering}
\caption{Amplitude transmission $\left|H(z,\omega\right|$ of the medium as
a function of the frequency modulus $\lvert\omega\rvert$ (logarithmic
scale). Parameters (in $\omega_{0}$ units) : $\omega_{p}=1.11$ ,
$\gamma=0.0707$ and $\xi=8.31\times10^{5}$ for the curve (a) corresponding
to the Brillouin choice ($z=10^{-2}\mathrm{m}$ ). The curves (b),
(c),(d) and (e) are obtained for propagation distances (and thus $\xi$
) respectively 10, 100, 1 000 and 10 000 times smaller.\label{fig:GainCurve}}
\end{figure}

Figure \ref{fig:GainCurve} shows the profiles of the amplitude transmission
$\left|H(z,\omega)\right|)=\exp\left[-\alpha\left(\omega\right)z\right]$
as a function of the reduced frequency $\omega/\omega_{0}$ in the
Brillouin conditions (curve a) and for propagation distances 10, 100,
1 000 and 10 000 times shorter (curves b to e).

The medium being opaque for $\omega_{-}<\omega<\omega_{+}$, the transfer
function may be written as 
\begin{equation}
H(z,\omega)=H_{S}(z,\omega)+H_{B}(z,\omega),\label{eq:six}
\end{equation}
 with $H_{S}(z,\omega)\approx0$ for $\omega<\omega_{+}$ and $H_{B}(z,\omega)\approx0$
for $\omega>\omega_{-}$. $H_{S}$ and $H_{B}$ are respectively associated
with the Sommerfeld and the Brillouin precursor. For $\omega=0$,
$H_{S}(z,0)\approx0$ and $H_{B}(z,0)\approx H(z,0)=1$. As long as
$\omega_{c}$ lies in the opacity region, this implies that \emph{the
Sommerfeld precursor will have a zero area while the area of the Brillouin
precursor will be equal to that of the incident field}.

The formation of the optical precursors is generally governed by combined
effects of attenuation (considered above) and dispersion. The dispersion
effects can be soundly characterized by the group delay $\tau_{g}(z,\omega)=-d\Phi/d\omega=z/v_{g}(\omega)-z/c$,
where $\Phi(z,\omega)$ is the argument of $H(z,\omega)$ and $v_{g}(\omega)$
the group velocity \cite{re2}. We remark that the regions of anomalous
dispersion ($dn/d\omega<0$) or of superluminal group velocity ($\tau_{g}<0$
) has a width smaller than $\omega_{0}$ and are entirely comprised
inside the opacity region. The corresponding frequencies will thus
not directly contribute to the formation of precursors. For the high
and low frequencies respectively associated with the Sommerfeld and
Brillouin precursors, we get the asymptotic forms $\tau_{g}\approx\xi/\omega^{2}$
\cite{re2} and $\tau_{g}\approx t_{B}+\omega^{2}/(\eta b^{3})$ where
\begin{equation}
t_{B}=\frac{\left[n(0)-1\right]z}{c}=\frac{2\xi}{\omega_{p}^{2}}\left[\left(1+\frac{\omega_{p}^{2}}{\omega_{0}^{2}}\right)^{1/2}-1\right]\label{eq:sept}
\end{equation}
 
\begin{equation}
b=\omega_{0}\left(3\frac{\xi}{\omega_{0}}\right)^{-1/3}\left(1+\frac{\omega_{p}^{2}}{\omega_{0}^{2}}\right)^{1/6}\label{eq:huit}
\end{equation}
 
\begin{equation}
\frac{1}{\eta}=1-\frac{4\gamma^{2}}{\omega_{0}^{2}}\left(1+\frac{3\omega_{p}^{2}}{4\omega_{0}^{2}}\right)/\left(1+\frac{\omega_{p}^{2}}{\omega_{0}^{2}}\right).\label{eq:neuf}
\end{equation}
 $t_{B}=\tau_{g}(z,0)-\tau_{g}(z,\infty)$ is obviously indicative
of the time delay of the Brillouin precursor (low frequency) with
respect to the Sommerfeld precursor (high frequency). The two precursors
will be fully separated when $t_{B}$ is much larger than the damping
time $1/\gamma$. Since $\gamma t_{B}=O\left(\gamma\xi/\omega_{0}^{2}\right)$,
this condition is automatically fulfilled when the condition of broad
opacity-region $\left[\gamma\xi/(10\omega_{0}^{2})\gg1\right]$ holds.
Another important point is that $\tau_{g}$ is minimum (stationary)
for $\omega\rightarrow\infty$ and $\omega\rightarrow0$. As pointed
out by Brillouin \cite{bri32}, this ensures that the precursors will
not be washed out by the group velocity dispersion.

\section{SOMMERFELD PRECURSOR\label{sec:SOMMERFELD-PRECURSOR} }

\subsection{Transfer function $H_{S}(z,\omega)$ and impulse response }

In the limit considered here $\omega^{2}\geq\omega_{+}^{2}\gg\omega_{0}^{2}$
and $H_{S}(z,\omega)$ takes the following asymptotic form, accounting
for both dispersion (main contribution) and attenuation. 
\begin{equation}
H_{S}(z,\omega)\approx\exp\left[-\frac{\xi}{i\omega+2\gamma}\right].\label{eq:dix}
\end{equation}
 The corresponding impulse response $h_{S}(z,t)$ is easily determined
by using standard results of Laplace transforms \cite{ab72}. We get
\begin{equation}
h_{S}(z,t)=\delta(t)-\sqrt{\frac{\xi}{t}}\mathrm{J}_{1}\left(2\sqrt{\xi t}\right)\mathrm{e}^{-2\gamma t}\mathrm{u}_{H}(t),\label{eq:onze}
\end{equation}
 where $\mathrm{J}_{n}(s)$ and $\mathrm{u}_{H}(t)$ respectively
designate the first kind Bessel-function of index $n$ and the Heaviside
unit-step function. Except for their very first oscillation, the Bessel
functions $\mathrm{J}_{n}(s)$ are perfectly approximated by their
asymptotic form 
\begin{equation}
\mathrm{J}_{n}(s)\approx\sqrt{\frac{2}{\pi s}}\cos\left(s-n\frac{\pi}{2}-\frac{\pi}{4}\right),\label{eq:douze}
\end{equation}
and the impulse response $h_{S}(z,t)$ can be characterized by an
instantaneous frequency $\omega\approx d\left(2\sqrt{\xi t}\right)/dt=\sqrt{\xi/t}$.
The range of validity of Eq.(\ref{eq:onze}) may be estimated by determining
the change $\delta H_{S}(z,\omega)$ of $H_{S}(z,\omega)$ due to
the first term neglected in the asymptotic expansion of $\ln\left[H_{S}\left(z,\omega\right)\right]$
used to obtain Eq.(\ref{eq:dix}). We find $\delta H_{S}(z,\omega)/H_{S}(z,\omega)=O\left(\xi\omega_{0}^{2}/\omega^{3}\right)$,
negligible when $\omega^{3}\gg\xi\omega_{0}^{2}$, i.e. when $\xi^{1/2}\gg\omega_{0}^{2}t^{3/2}$.
In fact, Eq.(\ref{eq:onze}) fits very well the exact impulse response
as soon as $\xi^{1/2}$ exceeds $\omega_{0}^{2}t^{3/2}$ by a factor
$\sqrt{10}$ (half an order of magnitude). This is achieved as long
as $t\leq t_{S}$, with 
\begin{equation}
\omega_{0}t_{S}=\sqrt[3]{\frac{\xi}{10\omega_{0}}}.\label{eq:treize}
\end{equation}
 In a strict asymptotic limit ($z\rightarrow\infty$), $t_{S}\rightarrow\infty$
and $\exp\left(-2\gamma t_{S}\right)\rightarrow0$. As expected, the
entirety of the impulse response is then reproduced by Eq.(\ref{eq:onze}).

\subsection{Precursor originated by a causal incident field}

The Sommerfeld precursor $e_{S}(z,t)$ is obtained by convoluting
$h_{S}(z,t)$ with the incident field $e(0,t)=u(t)\cos\left(\omega_{c}t-\varphi\right)$
introduced in the general analysis {[}Eq.(\ref{eq:zero}){]}. We are
mainly interested here in the physical case where the incident field
is causal {[}$e(0,t)=0$ for $t<0${]}, $u(t)$ being either a unit
step $\mathrm{u}_{H}(t)$ or a function \emph{monotonously} rising
from $0$ to $1$ with a rate $r\lesssim\omega_{c}$ for $t>0$ (step
or step-like modulation). The convolution product of Eq.(\ref{eq:quatre})
takes the form: 
\begin{equation}
e_{S}(z,t)=\intop_{-\infty}^{t}h_{S}(z,\theta)e(0,t-\theta)d\theta,\label{eq:quatorze}
\end{equation}
 that can be transformed by repeated integrations per parts to yield
\begin{equation}
e_{S}(z,t)=\sum_{n=0}^{\infty}d_{n}h_{S}^{\left(n+1\right)}(z,t).\label{eq:quinze}
\end{equation}
 Here $d_{n}$ is the discontinuity of the $\mathrm{n}^{\mathrm{th}}$
derivative of $e(0,t)$ at the initial time \cite{re3} and $f^{\left(n\right)}(t)$
is a short-hand notation for $\int_{-\infty}^{t}\int_{-\infty}^{t_{1}}\cdots\int_{-\infty}^{t_{n-1}}f(t_{n})dt_{n}\cdots dt_{2}dt_{1}$.
In a frequency description, the previous result can be retrieved by
expanding the Fourier transform $E(0,\omega)$ of $e(0,t)$ in powers
of $1/i\omega$ and exploiting the equivalence between multiplication
by $1/i\omega$ in the frequency domain and integration in the time
domain \cite{pap87}. Writing the impulse response under the form
$h_{S}(z,t)=k_{S}(z,t)\exp(-2\gamma t)$, we easily show by means
of standard Laplace procedures \cite{ab72} that $k_{S}^{\left(n+1\right)}(z,t)=(t/\xi)^{n/2}\mathrm{J}_{n}(2\sqrt{\xi t})\mathrm{u}_{H}(t)$.
Insofar as $k_{S}(z,t)$ is very rapidly varying compared to $\exp(-2\gamma t)$,
$h_{S}^{\left(n+1\right)}(z,t)\approx k_{S}^{\left(n+1\right)}(z,t)\exp(-2\gamma t)$
and we finally get 
\begin{equation}
e_{S}(z,t)\approx\sum_{n=0}^{\infty}d_{n}\left(\frac{t}{\xi}\right)^{n/2}\mathrm{J}_{n}(2\sqrt{\xi t})\exp\left(-2\gamma t\right)\mathrm{u}_{H}(t).\label{eq:seize}
\end{equation}
 The $\mathrm{n^{th}}$ term of the series has a maximal amplitude
$a_{0}=\left|d_{0}\right|$ at $t=t_{0}=0$ for $n=0$ and 
\begin{equation}
a_{n}=\frac{1}{\sqrt{\pi}}\left|d_{n}\right|\left(\frac{2n-1}{8\mathrm{e}}\right)^{\left(2n-1\right)/4}\left(\frac{\gamma}{\xi}\right)^{1/4}\left(\gamma\xi\right)^{-n/2},\label{eq:dixsept}
\end{equation}
 at $t\approx t_{n}=\left(2n-1\right)/8\gamma$ for $n>0$. Since
$\xi\propto z$, Eq.(\ref{eq:dixsept}) shows that, for large propagation
distance, $a_{n}$ rapidly decreases with $n$, so that a good approximation
of the exact result is obtained by keeping only the first term $n=p$
of the series for which $d_{p}\neq0$. In the frequency description,
this amounts to restrict the asymptotic expansion of $E(0,\omega)$
to its first non vanishing term \cite{ja75}. We then get 
\begin{equation}
e_{S}(z,t)\approx d_{p}\left(\frac{t}{\xi}\right)^{p/2}\mathrm{J}_{p}(2\sqrt{\xi t})\exp\left(-2\gamma t\right)\mathrm{u}_{H}(t).\label{eq:dixhuit}
\end{equation}
 Denoting $q$ is the next integer following $p$ for which $d_{q}\neq0$,
Eq.(\ref{eq:dixhuit}) is exact when $\varepsilon=a_{q}/a_{p}\approx0$,
and $\exp\left(-2\gamma t_{S}\right)\approx0$. These conditions are
met in the strict asymptotic limit and closely approached for the
propagation distance considered by Brillouin. At distances that may
be 1 000 times smaller (simple asymptotic limit); we shall see that
Eq.(\ref{eq:dixhuit}) enables us to correctly reproduce the essential
features of the precursor originated by representative incident fields.

\subsection{Precursor originated by a discontinuous incident field}

We consider first the instructive case where $e(0,t)=u_{H}(t)\cos\left(\omega_{c}t\right)$
for which $p=0$ with $d_{0}=1$ \cite{re3} and $q=2$ with $d_{2}=-\omega_{c}^{2}$.
Eq.(\ref{eq:dixhuit}) then reads as 
\begin{equation}
e_{S}(z,t)\approx\mathrm{J_{0}}(2\sqrt{\xi t})\exp\left(-2\gamma t\right)\mathrm{u}_{H}(t),\label{eq:dixneuf}
\end{equation}
 with $\varepsilon\approx0.13\omega_{c}^{2}\gamma^{-3/4}\xi^{-5/4}$
{[}see Eq. \ref{eq:dixsept}{]}. The precursor does not depend on
$\omega_{c}$ and the initial discontinuity of the incident field
is integrally transmitted, in agreement with the general analysis.
For $\omega_{c}<\omega_{+}=\sqrt{\gamma\xi/10}$ (opacity condition),
$\varepsilon$ is always smaller than $0.013\left(\gamma/\xi\right)^{1/4}$,
that is about $2.2\times10^{-4}$ in the Brillouin conditions and
$1.2\times10^{-3}$ for a propagation distance 1 000 times smaller
(simple asymptotic limit). In the first case, $\omega_{0}t_{S}=44$
and $\exp\left(-2\gamma t_{S}\right)\approx2\times10^{-3}$. As previously
indicated, we are then close to the strict asymptotic limit and the
precursor is perfectly reproduced by its asymptotic form at any time
where it has a significant amplitude. 
\begin{figure}[h]
\begin{centering}
\includegraphics[width=80mm]{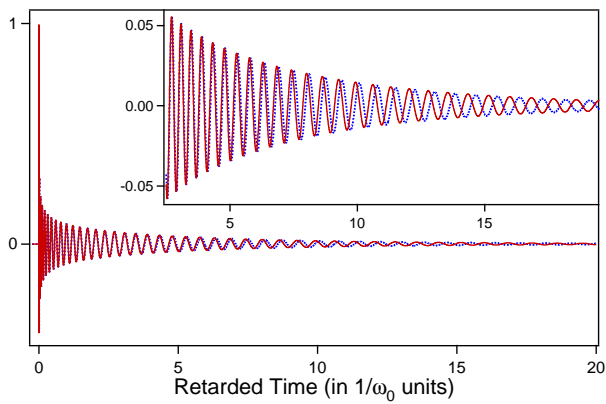} 
\par\end{centering}

\caption{Sommerfeld precursor originated by the incident field $\cos(\omega_{c}t)\mathrm{u}_{H}(t)$.
The solid (dashed) line is the exact numerical solution (the approximate
analytic solution). Parameters (in $\omega_{0}$ units) : $\omega_{c}=1$,
$\omega_{p}=1.11$, $\gamma=0.0707$ and $\xi=831$. Inset : enlargement
of the tail of the precursor. \label{fig:SommerfeldCondBrCos}}
\end{figure}

This remark also holds for the cases considered in the following subsections.
In the simple asymptotic limit $\omega_{0}t_{S}=4.4$ and, as expected,
Eq.(\ref{eq:dixneuf}) perfectly fits the exact solution for $\omega_{0}t\leq4.4$.
For larger times, the fit remains very good except for a slight drift
of the instantaneous frequency of the oscillations whose envelope
is very well reproduced at any time (Fig.\ref{fig:SommerfeldCondBrCos}).

\subsection{Precursor originated by the canonical incident field of Sommerfeld
and Brillouin}

\begin{figure}[h]
\begin{centering}
\includegraphics[width=80mm]{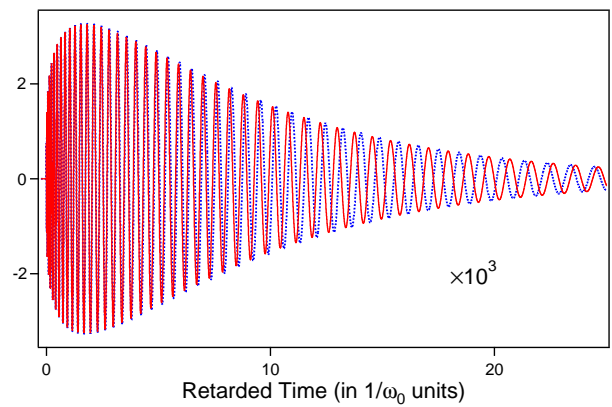} 
\par\end{centering}

\caption{Sommerfeld precursor originated by the canonical incident field $\sin(\omega_{c}t)\mathrm{u}_{H}(t)$.
The solid (dashed) line is the exact numerical solution (the approximate
analytic solution). Parameters as in Fig.\ref{fig:SommerfeldCondBrCos}.
\label{fig:Sommerfeld_sin}}
\end{figure}

Following Sommerfeld and Brillouin, most authors have considered an
incident field of the \emph{canonical form} $e(0,t)=u_{H}(t)\sin\left(\omega_{c}t\right)$
for which $p=1$ with $d_{1}=\omega_{c}$ and $q=3$ with $d_{3}=-\omega_{c}^{3}$.
We then get 
\begin{equation}
e_{S}(z,t)\approx\omega_{c}\sqrt{\frac{t}{\xi}}\mathrm{J_{1}}(2\sqrt{\xi t})\exp\left(-2\gamma t\right)\mathrm{u}_{H}(t),\label{eq:vingt}
\end{equation}
 with $\varepsilon\approx0.34\left(\omega_{c}^{2}/\gamma\xi\right)$.
The result given Eq.(\ref{eq:vingt}) differs from that originally
obtained by Sommerfeld \cite{som14} by the presence of the damping
term $\exp\left(-2\gamma t\right)$ . Though \emph{the formation of
the Sommerfeld precursor is mainly governed by the medium dispersion,}
the presence of this term (associated with the absorption) is obviously
necessary to avoid that $e_{S}(z,t)$ diverges with time. The precursor
attains its maximum at $t\approx t_{1}=1/\left(8\gamma\right)$ ($\omega_{0}t_{1}=1.77$
) and its amplitude $a_{S}=a_{1}\approx0.26\:\omega_{c}\gamma^{-1/4}\xi^{-3/4}$
is \emph{proportional to} $\omega_{c}$. For $\omega_{c}=\omega_{0}$,
$a_{S}\approx1.8\times10^{-5}$ with $\varepsilon\approx5.8\times10^{-6}$
in the Brillouin conditions whereas $a_{S}\approx3.25\times10^{-3}$
with $\varepsilon\approx5.8\times10^{-3}$ in the simple asymptotic
limit. In the latter case, Fig.\ref{fig:Sommerfeld_sin} shows that
Eq.(\ref{eq:vingt}) actually fits very well the exact result for
$t\leq t_{S}$, again with a slight drift of the instantaneous frequency
of the oscillations for $t>t_{S}$. In order to check the proportionality
of the precursor to $\omega_{c}$, we have compared the exact forms
of $\left(\omega_{0}/\omega_{c}\right)\: e_{S}(z,t)$ obtained when
$\omega_{c}$ lies at the boundaries $\omega_{-}$ or $\omega_{+}$
of the opacity region to that obtained when $\omega_{c}=\omega_{0}$.
As expected we have found that the three results are nearly undistinguishable,
except for an amplitude $1.3\%$ larger for $\omega_{c}=\omega_{+}$
(below the corresponding value of $\epsilon$, namely $\varepsilon=0.034$).
For this value of $\omega_{c}$, the amplitude of the precursor is
$a_{S}\approx0.082\left(\gamma/\xi\right)^{1/4}$, that is $1.40\times10^{-3}$
in the Brillouin conditions and $7.9\times10^{-3}$ in the simple
asymptotic limit.

\subsection{Rise-time effects }

\begin{figure}[h]
\begin{centering}
\includegraphics[width=80mm]{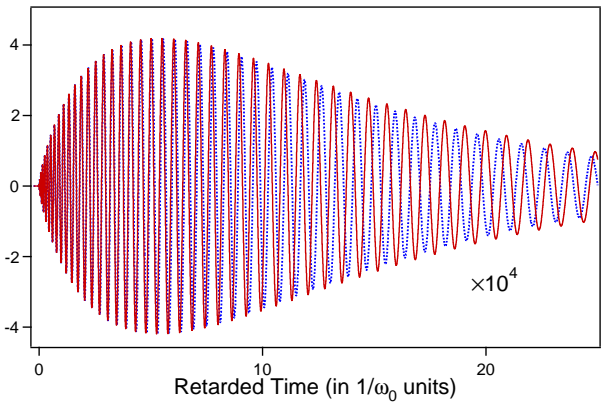} 
\par\end{centering}

\caption{Sommerfeld precursor originated by the incident field $e(0,t)=\tanh(rt)\:\sin(\omega_{c}t)\: u_{H}(t)$.
The solid (dashed) line is the exact numerical solution (the approximate
analytic solution) obtained for $r=\omega_{0}$. Other parameters
as in Fig.\ref{fig:SommerfeldCondBrCos}\label{fig:SommerfeldRiseTime}.}
\end{figure}

A gradual turning on of the incident field is expected to reduce the
amplitude of the Sommerfeld precursor. To study this so-called rise-time
effect, Ciarkowski \cite{cia02,cia11} has considered the incident
field $e(0,t)=\tanh\left(rt\right)\sin\left(\omega_{c}t\right)\mathrm{u}_{H}(t)$
whose envelope has a $10-90\%$ rise time $T_{r}\approx1.37/r$. In
this case $p=2$ with $d_{2}=2r\omega_{c}$, $q=4$ with $d_{4}=-4\omega_{c}r\left(2r^{2}+\omega_{c}^{2}\right)$
and the asymptotic form of the precursor reads as 
\begin{equation}
e_{S}(z,t)\approx2\omega_{c}r\left(\frac{t}{\xi}\right)\mathrm{J}_{2}(2\sqrt{\xi t})\exp\left(-2\gamma t\right)\mathrm{u}_{H}(t),\label{eq:vingtetun}
\end{equation}
 with $\varepsilon\approx1.21\left(2r^{2}+\omega_{c}^{2}\right)/\gamma\xi$.
The precursor attains its maximum at $t\approx t_{2}=3/(8\gamma)$
($\omega_{0}t_{2}\approx5.3$) with an amplitude $a_{S}=a_{2}\approx0.26\: r\omega_{c}\gamma^{-3/4}\xi^{-5/4}$.
Compared to the precursor obtained with the canonical incident field
{[}Eq.(\ref{eq:vingt}){]}, the maximum is shifted to larger time
($t_{2}=3t_{1}$) and its amplitude is reduced by a factor $\rho\approx\sqrt{\gamma\xi}$/r.
Fig.\ref{fig:SommerfeldRiseTime}, obtained in the simple asymptotic
limit, shows that Eq.(\ref{eq:vingtetun}) fits quite satisfactorily
the exact precursor though its maximum now lies at a time slightly
larger than $t_{S}$. To check that the precursor is mainly determined
by the lowest order initial discontinuity of the incident field regardless
of its subsequent evolution, we have compared the precursor obtained
when the envelope $\tanh\left(rt\right)u_{H}(t)$ is replaced by $\left(1-\mathrm{e}^{-rt}\right)u_{H}(t)$,
having the same initial discontinuity. Though $q=3$ (instead of $4$)
and $T_{r}\approx2.20/r$ (instead of $1.37/r$), we have found that
the precursor is actually very close to the previous one.

Other things being equal, the reduction of the amplitude of the precursor
is more and more important when the incident field is applied more
and more smoothly, that is when the order $p$ of its initial discontinuity
increases. It is easily deduced from Eq.(\ref{eq:dixsept}) that for
$p\geq2$, $\rho=O\left[\left(\sqrt{\gamma\xi}/r\right)^{p-1}\right]\propto\left(T_{r}\sqrt{z}\right)^{p-1}$.
At the light of this result, dramatic rise time effects are expected
when the incident field is ideally smooth, i.e. analytic with continuous
derivatives in every point. Such fields have been considered \cite{ou09,bm11,ou95}
though they are not causal and, strictly speaking, not physically
realizable (in the sense of the linear systems theory). We have made
numerical simulations for $e(0,t)=\sin\left(\omega_{c}t\right)\left[1+\mathrm{erf}\left(rt\right)\right]/2$
where $\mathrm{erf}(s)$ designates the error function. For $z$ and
$r=\omega_{0}$ as in Fig.4, we get $\rho\approx1.4\times10^{3}$
instead of $\rho\approx7.7$ for $e(0,t)=\tanh(rt)\:\sin(\omega_{c}t)\: u_{H}(t)$. 

\section{BRILLOUIN PRECURSOR IN THE STRICT ASYMPTOTIC LIMIT\label{sec:BrillouinStrictLimit} }

\subsection{Transfer function $H_{B}(z,\omega)$ and impulse response}

In the limit considered now $\omega^{2}\leq\omega_{-}^{2}\ll\omega_{0}^{2}$
and $H_{B}(z,\omega)$ is conveniently developed under the form 
\begin{equation}
H_{B}(z,\omega)=\exp\left(\sum_{n=1}^{\infty}\frac{\left(-i\omega\right)^{n}}{n!}k_{n}(z)\right).\label{eq:vingttrois}
\end{equation}
 Here $k_{n}(z)$ are the so-called cumulants, generally introduced
in probability theory \cite{ab72}, but also quite useful to study
deterministic signals \cite{bu04,bm06}. The cumulants $k_{1}(z)$,
$k_{2}(z)$ and $k_{3}(z)$ have remarkable properties. $k_{1}(z)$
and $k_{2}^{1/2}(z)$ respectively are the center-of-mass and the
root-mean-square duration of the impulse response $h_{B}(z,t)$, inverse
Fourier transform of $H_{B}(z,\omega)$, whereas $\kappa(z)=k_{3}(z)/k_{2}^{3/2}(z)$
is its normalized asymmetry or skewness \cite{ab72}. From Eqs.(\ref{eq:deux},\ref{eq:trois}),
we easily get $k_{1}=t_{B}$ (as expected), $k_{2}=4\gamma/(3b^{3})$
, $k_{3}=-2/(\eta b^{3})$ and $\kappa=-(1/4\eta)\left(3b/\gamma\right)^{3/2}$,
where $t_{B}$, $b$ and $\eta$ are defined by Eqs.(\ref{eq:sept}-\ref{eq:neuf}).
When $z\rightarrow\infty$ (strict asymptotic limit), $\kappa\propto b^{3/2}\propto z^{-1/2}\rightarrow0$
and the expansion of Eq.(\ref{eq:vingttrois}) may be limited to the
term $n=2$. Taking a new origin of time at $t=t_{B}$, the transfer
function then reads as 
\begin{equation}
H_{B}(z,\omega)\approx\exp\left(-\frac{\omega^{2}}{4\beta^{2}}\right),\label{eq:vingtquatre}
\end{equation}
 where $\beta=\sqrt{3b^{3}/8\gamma}\propto1/\sqrt{z}$ is very small
compared to $\omega_{0}$. This Gaussian form is that of the normal
distribution derived by means of the central limit theorem in probability
theory. This theorem can also be used to obtain an approximate evaluation
of the convolution of $n$ deterministic functions \cite{pap87}.
It can be applied to our case by splitting the medium into $n$ cascaded
sections, $h_{B}(z,t)$ being the convolution of the impulses responses
of each section. By calculating the inverse Fourier transform of $H_{B}(z,\omega)$,
we get 
\begin{equation}
h_{B}(z,t)=\frac{\beta}{\sqrt{\pi}}\exp\left(-\beta^{2}t'^{2}\right),\label{eq:vingtcinq}
\end{equation}
 where $t'=t-t_{B}$. The impulse response has a duration (amplitude)
proportional (inversely proportional) to $\sqrt{z}$, with an area
constantly equal to $1$ (in agreement with the general analysis).
We remark that the approximation leading to Eq.(\ref{eq:vingtquatre})
and Eq.(\ref{eq:vingtcinq}), valid in the strict asymptotic limit,
amounts to neglect the effects of the group delay dispersion, the
formation of the Brillouin precursor being then governed by the frequency
dependence of the medium attenuation (dominant-attenuation limit).

The Gaussian forms of Eq.(\ref{eq:vingtquatre}) and Eq.(\ref{eq:vingtcinq})
are not specific to the Lorentz medium but have some generality \cite{kla05}.
They hold for the Debye medium \cite{sto01}, for some random media
\cite{ga10} and, more generally, whenever the transfer function of
the medium can be expanded in cumulants and the propagation distance
is such that $\left|\kappa\right|\ll1$. Stoudt \emph{et al}. \cite{sto01}
showed in particular that the results of their experiments on water
(Debye medium) at decimetric wavelengths can be numerically reproduced
by neglecting the group delay dispersion, as it has been made to obtain
Eq.(\ref{eq:vingtquatre}). See also \cite{ou05,pi09,da10,ca11}.
Using a purely temporal approach, Karlsson and Ritke \cite{ka98}
early remarked that the impulse response of the Debye medium is very
close to a normalized Gaussian. This property is obviously a consequence
of the previous analysis. The complex refractive index now reads as
$\widetilde{n}(\omega)=\left[1+\left(n_{0}^{2}-1\right)/\left(1+i\omega\tau\right)\right]^{1/2}$
where $n_{0}$ is the refractive index for $\omega\rightarrow0$ and
$\tau$ is the relaxation time for the orientation of the polar molecules
\cite{pi09}. Including $\widetilde{n}(\omega)$ in Eq.(\ref{eq:deux})
and following the procedure used for the Lorentz medium, we easily
get $\beta=\left[2\left(n_{0}^{2}-1\right)\tau z/cn_{0}\right]^{-1/2}$
and, taking into account that $n_{0}^{2}\gg1$, $\kappa\approx2.25\sqrt{c\tau/n_{0}z}$.
Note that $\beta$ and $\kappa$ depends on $z$ as $1/\sqrt{z}$
(as in the Lorentz medium). The normalized Gaussian of Eq.(\ref{eq:vingtcinq})
will thus also be obtained for sufficient propagation distances. Using
the parameters of water \cite{pi09}, namely $n_{0}=\sqrt{79}$ and
$\tau=8.5\times10^{-12}\mathrm{s}$, we find that the skewness of
$5.2\%$, obtained in a Lorentz medium for a propagation distance
larger by more of four orders of magnitude than the optical wavelengths
considered, is now attained for a propagation distance $z\approx0.55\:\mathrm{m}$
comparable to the wavelengths involved in the experiments reported
in \cite{sto01}. Despite strongly different scales, Brillouin precursors
in the Lorentz medium in the strict asymptotic limit and in the Debye
medium pertain to the same physics, namely that of the dominant-attenuation
limit, and will be described by the same laws. On the other hand,
the Debye medium is fully opaque at high frequency and Sommerfeld
precursors cannot be generated in this medium.

\subsection{Precursor generated by an incident field of non-zero area}

The Brillouin precursor generated by an arbitrary incident field $e(0,t)$
is obtained by convoluting the latter with $h_{B}(z,t)$ or by multiplying
its Fourier transform $E(0,\omega)$ by $H_{B}(z,\omega)$ and determining
the inverse Fourier transform of the product. We consider first the
case where $e(0,t)$ is rapidly varying compared to $h_{B}(z,t)$.
This requires in particular that $\omega_{c}\gg\beta$. Compared to
$E(0,\omega)$, $H_{B}(z,\omega)$ then appears as a narrow peak centered
on $\omega=0$ and, provided that $E(0,0)\neq0$, $E_{B}(z,\omega)\approx E(0,0)\: H_{B}(z,\omega)$.
Remembering that $E(0,0)$ is the algebraic area $\mathcal{A}$ of
the incident field (see Sec. \ref{sec:GENERAL-ANALYSIS}), we finally
get: 
\begin{equation}
e_{B}(z,t)\approx\mathcal{A}h_{B}(z,t)=\frac{\mathcal{A\beta}}{\sqrt{\pi}}\exp\left(-\beta^{2}t'^{2}\right).\label{eq:vingtsix}
\end{equation}
 For the canonical incident field $\sin\left(\omega_{c}t\right)\mathrm{u}_{H}(t)$,
$E(0,0)=1/\omega_{c}$ and the precursor has an amplitude $a_{B}=\beta/\left(\omega_{c}\sqrt{\pi}\right)$
\emph{inversely proportional} to $\omega_{c}$ (no matter its value
provided that $\omega_{c}\gg\beta$) and to $\sqrt{z}$. Note that
the law $a_{B}\propto1/\sqrt{z}$, sometimes considered as general,
is only valid in the strict asymptotic limit considered here (for
which $\left|\kappa\right|\ll1$). 
\begin{figure}[h]
\begin{centering}
\includegraphics[width=80mm]{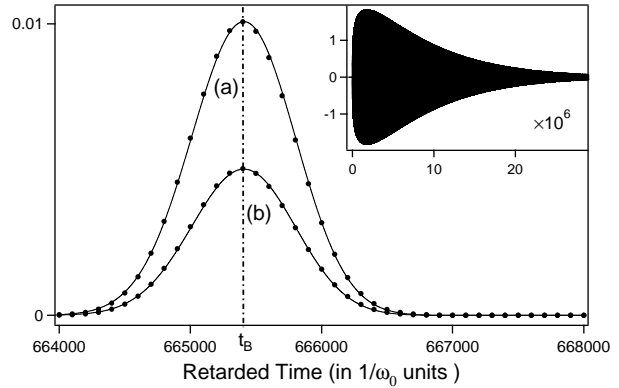} 
\par\end{centering}

\caption{Brillouin precursor obtained in the Brillouin conditions, namely for
$\omega_{c}=0.1$, $\omega_{p}=1.11$, $\gamma=0.0707$ and $\xi=8.31\times10^{5}$
(in $\omega_{0}$ units). For these parameters, $\omega_{0}t_{B}\approx6.654\times10^{5}$
and $\beta\approx1.78\times10^{-3}\omega_{0}=1.78\times10^{-2}\omega_{c}$.
Solid lines (bullets $\bullet$) are the exact numerical solutions
(the analytic solutions). Curve (a) is the precursor obtained with
the canonical incident field $\sin(\omega_{c}t)u_{H}(t)$. The precursor
of curve (b) is originated by the incident field $e(0,t)=\sin(\omega_{c}t)\left[1+\textrm{erf}(rt)\right]/2$
for $r=\omega_{c}/2\sqrt{2}$ . Inset: Sommerfeld precursor obtained
in the conditions of curve (a). It fully vanishes in the conditions
of curve (b). \label{fig:BrillouinVraiSin}}
\end{figure}
Fig.\ref{fig:BrillouinVraiSin} shows that the precursor obtained
in \emph{all} the Brillouin conditions {[}curve (a){]} is perfectly
fitted by the Gaussian form of Eq.(\ref{eq:vingtsix}). We incidentally
note that, for the carrier frequency retained by Brillouin ($\omega_{c}=\omega_{0}/10$),
the medium is fully opaque at this frequency {[}$\alpha\left(\omega_{c}\right)z\approx800${]},
in contradiction with his artist's view showing a ``main field\textquotedblright{}
(at $\omega_{c}$) larger than the precursors. On the other hand,
the condition $\omega_{c}\gg\beta$ is well satisfied. The inset in
Fig.\ref{fig:BrillouinVraiSin} shows the Sommerfeld precursor obtained
in the same conditions. As already mentioned, it is perfectly fitted
by the analytical expression of Eq.(\ref{eq:vingt}). Note however
that its amplitude is about four orders of magnitude smaller than
that of the Brillouin precursor. Eq.(\ref{eq:vingtsix}) also holds
when the envelope of the incident field rises in a finite time provided
that the rate $r$ , as $\omega_{c}$, is large compared to $\beta$.
Curve (b) of Fig.\ref{fig:BrillouinVraiSin} shows the Brillouin precursor
generated by the incident field $e(0,t)=\sin\left(\omega_{c}t\right)\left[1+\mathrm{erf}\left(rt\right)\right]/2$.
We have then $E(0,0)=(1/\omega_{c})\exp\left(-\omega_{c}^{2}/4r^{2}\right)$
and the area of the incident pulse, equal to $1/\omega_{c}$ for $r\rightarrow\infty$,
falls to $1/2\omega_{c}$ for $r=\omega_{c}/2\sqrt{\ln\left(2\right)}$
($r\approx0.60\omega_{c}$). As expected, the Brillouin precursor
is identical to the previous one with amplitude reduced by half and
the corresponding Sommerfeld precursor completely vanishes.

\subsection{Precursor originated by an incident field of zero area}

\begin{figure}[h]
\begin{centering}
\includegraphics[width=80mm]{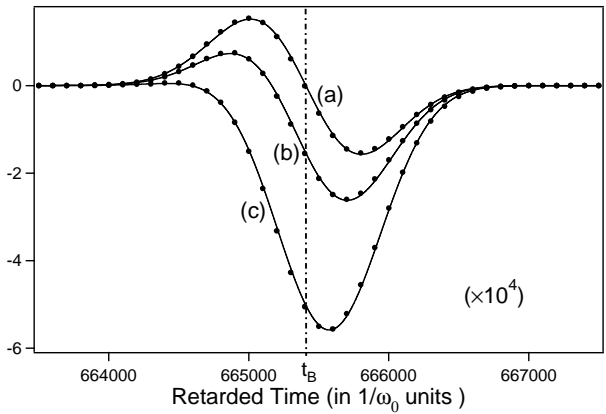} 
\par\end{centering}

\caption{Brillouin precursor obtained with the incident fields $(1-\mathrm{e}^{-rt})\cos(\omega_{c}t)\mathrm{u}_{H}(t)$
for (a) $r\rightarrow\infty$, (b) $r=65\omega_{c}$ and (c) $r=20\omega_{c}$
(solid lines). Other parameters as in Fig.\ref{fig:BrillouinVraiSin}.
The bullets $\bullet$ correspond to the analytical solutions given
by Eq.(\ref{eq:vingtsept}) or by the combination of this equation
with Eq.(\ref{eq:vingtsix}). \label{fig:BrillouinVraiCos}}
\end{figure}

Even if $\omega_{c},r\gg\beta$, Eq.(\ref{eq:vingtsix}) obviously
fails when $\mathcal{A}=E(0,0)=0$. This occurs in particular in the
extreme case where the incident field is instantaneous turned on,
with $e(0,t)=\cos\left(\omega_{c}t\right)\mathrm{u}_{H}(t)$. It is
then necessary to consider the next term in the expansion of $E(0,\omega)$
in powers of $i\omega$. We get in this case $E(0,\omega)\approx i\omega/\omega_{c}^{2}$
and $E_{B}(z,\omega)\approx i\omega H_{B}(z,\omega)/\omega_{c}^{2}$.
Using the correspondence $i\omega\leftrightarrow d/dt$ between frequency
and time descriptions \cite{pap87} and denoting by a dot the time
derivative, we finally get: 
\begin{equation}
e_{B}(z,t)\approx\frac{1}{\omega_{c}^{2}}\overset{.}{h_{B}}(z,\omega t)=-\frac{2\beta^{2}}{\omega_{c}^{2}\sqrt{\pi}}\:\beta t'\exp\left(-\beta^{2}t'^{2}\right).\label{eq:vingtsept}
\end{equation}
 As shown Fig.\ref{fig:BrillouinVraiCos} {[}curve (a){]}, the analytical
expression of Eq.(\ref{eq:vingtsept}) perfectly fits the exact numerical
results obtained by FFT. The precursor is a Gaussian derivative with
a peak amplitude $a_{B}=\left[2/(\pi\mathrm{e})\right]^{1/2}\left(\beta/\omega_{c}\right)^{2}$,
smaller than that attained with the canonical incident field by a
factor $\omega_{c}\sqrt{\mathrm{e}}/(\beta\sqrt{2})$ ($\approx65$
in \emph{all} the Brillouin conditions) and decreasing much more rapidly
with the propagation distance (as $1/z$ instead of as $1/\sqrt{z}$).
We however remark that the precursor so obtained is not robust. Indeed
it suffices that the incident field suffers a short rise time to retrieve
a precursor mainly governed by the area law of Eq.(\ref{eq:vingtsix}).
To illustrate this point, we have again considered an incident field
of the form $\left(1-\mathrm{e}^{-rt}\right)\cos\left(\omega_{c}t\right)\mathrm{u}_{H}(t)$
that tends to $\cos\left(\omega_{c}t\right)\mathrm{u}_{H}(t)$ for
$r\rightarrow\infty$. For $r\gg\omega_{c}$ (very short rise time),
$E(0,\omega)\approx-1/r+i\omega/\omega_{c}^{2}$. The incident field
has gained a (negative) area $\mathcal{A}=-1/r$. The precursor is
then the sum of two contributions, respectively given by Eq.(\ref{eq:vingtsix})
with $\mathcal{A}=-1/r$ and by Eq.(\ref{eq:vingtsept}). Curve (b)
of Fig.\ref{fig:BrillouinVraiCos} shows the result obtained when
the two contributions have the same amplitude, that is when $r/\omega_{c}=\omega_{c}\sqrt{\mathrm{e}}/(\beta\sqrt{2})\approx65$
. When $r$ decreases by remaining large compared to $\omega_{c}$,
the Gaussian part of the precursor rapidly prevails on the Gaussian-derivative
part and, as shows {[}curve (c){]}, the precursor becomes nearly Gaussian
(downwards) for $r$ as large as $20\omega_{c}$.

\subsection{Case where the carrier frequency lies below the opacity region}

\begin{figure}[h]
\begin{centering}
\includegraphics[width=80mm]{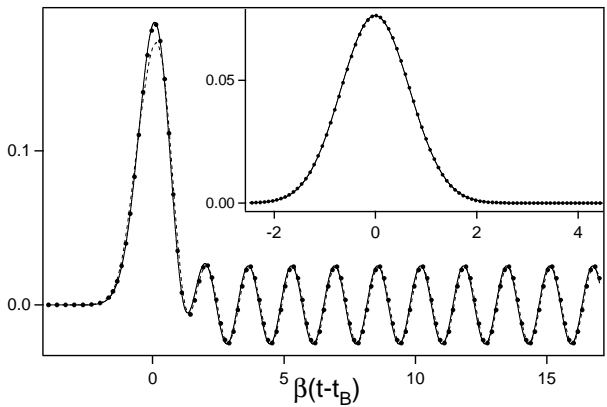} 
\par\end{centering}
\caption{Brillouin precursor and main field obtained for $\omega_{c}\approx3.84\beta$
as a function of $\beta(t-t_{B})$. The solid line, the bullets $\bullet$
and the dashed line are respectively the exact numerical solution,
the analytical solution given Eq.(\ref{eq:vingtneuf}) and its approximate
form given Eq.(\ref{eq:trente}). Inset: Brillouin precursor obtained
for $\omega_{c}\approx7.67\beta$. The two analytical solutions are
undistinguishable in this case and the amplitude of the main field
is negligible. \label{fig:BrillouinMainField}}
\end{figure}

The previous results are valid for the Lorentz medium in the strict
asymptotic limit (also as in the Debye medium) when $\omega_{c}\gg\beta$,
that is when $\omega_{c}$ lies in the opacity region. Fortunately
enough, the simplicity of the Gaussian impulse response enables us
to obtain exact expressions of the transmitted field for arbitrary
values of the ratio $\omega_{c}/\beta$. This occurs in the Lorentz
medium when $\omega_{c}$ resides below the opacity region and \emph{direct}
observations of the field transmitted in such conditions have been
performed by Stoudt \emph{et al}. in a Debye medium \cite{sto01}.
The transmitted field $e(z,t)$ is calculated directly in the time
domain by convoluting $h_{B}(z,t)$ given Eq.(\ref{eq:vingtcinq})
with the incident field. For the canonical incident field, the convolution
product can be written as: 
\begin{equation}
e(z,t)=\frac{\beta}{\sqrt{\pi}}\intop_{-\infty}^{t'}\mathrm{e}^{-\beta^{2}\theta^{2}}\sin\left[\omega_{c}\left(t'-\theta\right)\right]d\theta.\label{eq:vingthuit}
\end{equation}
 After some simple transformations, we finally get 
\begin{equation}
e(z,t)=\frac{1}{2}\:\mathrm{e}^{-\omega_{c}^{2}/4\beta^{2}}\Im\left\{ \left[1+\mathrm{erf}\left(\beta t'+\frac{i\omega}{2\beta}\right)\right]\:\mathrm{e}^{i\omega_{c}t'}\right\} \label{eq:vingtneuf}
\end{equation}
 where $\mathrm{e}^{-\omega_{c}^{2}/4\beta^{2}}\approx\mathrm{e}^{-\alpha\left(\omega_{c}\right)z}$
and, as previously, $t'=t-t_{B}$. For $t'\rightarrow\infty$, $e(z,t)$
tends to $\mathrm{e}^{-\omega_{c}^{2}/4\beta^{2}}\sin\left(\omega_{c}t'\right)$
which is nothing but that the steady state or main field, not negligible
when $\omega_{c}$ and $\beta$ are comparable. If we take $t_{B}$
($1/\beta$) as time origin (time unit), the transmitted field only
depends on the ratio $\omega_{c}/\beta$, \emph{regardless of the
particular system considered}. When $\omega_{c}\gg\beta$, it tends
to $\beta/\left(\omega_{c}\sqrt{\pi}\right)\exp\left(-\beta^{2}t'^{2}\right)$
in agreement with Eq.(\ref{eq:vingtsix}), the main field being then
negligible. When $\omega_{c}\geq4\beta$, Eq.(\ref{eq:vingtneuf})
is well approximated by the expression: 
\begin{multline}
e(z,t)\approx\frac{1+\mathrm{erf}\left(\beta t'\right)}{2}\:\sin\left(\omega_{c}t'\right)\:\mathrm{e}^{-\alpha\left(\omega_{c}\right)z}+\\
\frac{\beta'}{\omega_{c}\sqrt{\pi}}\:\mathrm{e}^{-\beta'^{2}t'^{2}},\label{eq:trente}
\end{multline}
 where $\beta'=\beta\left(1+2\beta^{2}/\omega_{c}^{2}\right)\rightarrow\beta$
for $\omega_{c}\gg\beta$. The first (second) term of Eq.(\ref{eq:trente})
obviously corresponds to the main field (the Brillouin precursor).
Figure \ref{fig:BrillouinMainField} shows the transmitted field as
a function of $\beta t'=\beta\left(t-t_{B}\right)$ for $\omega_{c}\approx3.84\beta$
and $\omega_{c}\approx7.67\beta$ (inset). In the study on water (Debye
medium) at decimetric wavelengths \cite{sto01}, these values are
obtained with $\omega_{c}=2\pi\times10^{9}\:\mathrm{s}^{-1}$, for
$z=0.75\:\mathrm{m}$ and $z=3\:\mathrm{m}$ respectively. As expected
Eq.(\ref{eq:vingtneuf}) perfectly fits the exact numerical result
in both cases. Eq.(\ref{eq:trente}) provides a good approximation
for $\omega_{c}\approx3.84\beta$, excellent for $\omega_{c}\approx7.67\beta$.
In the latter case, the Brillouin precursor prevails over the main
field whose relative amplitude is negligible. The signals shown Fig.\ref{fig:BrillouinMainField}
are in good agreement with those directly observed in the experiments
reported in \cite{sto01}.

\section{EXTENDED EXPRESSION OF THE BRILLOUIN PRECURSOR\label{sec:EXTENDED-EXPRESSION-BP}}

We come back in this section to the Brillouin precursor in the Lorentz
medium. Numerical simulations show that the solutions obtained in
the strict asymptotic or dominant-attenuation limit continue to provide
good (not too bad) approximations of the exact solutions when the
propagation distance is $10$ times ($100$ times) shorter than that
considered by Brillouin \cite{re1}, though the skewness $\kappa$
then rises up to $16\%$ ($52\%$). For shorter distances, it is obviously
necessary to take into account the effects of the group-delay dispersion
neglected in the strict asymptotic approximation.

\subsection{Transfer function $H_{B}(z,\omega)$ and impulse response}

Taking into account the term in $\omega^{3}$ in Eq. (\ref{eq:vingttrois}),
the transfer function then reads as 
\begin{equation}
H_{B}(z,\omega)\approx\exp\left[-i\omega t_{B}-\frac{i}{3\eta b^{3}}\left(\omega^{3}-2i\eta\gamma\omega^{2}\right)\right],\label{eq:trenteetun}
\end{equation}
 where $t_{B}$, $b$ and $\eta$ are defined by Eqs.(\ref{eq:sept}-\ref{eq:neuf}),
with $2\gamma/(3b^{3})=1/4\beta^{2}$. Remarking that $\left(\omega^{3}-2i\eta\gamma\omega^{2}\right)$
is the beginning of $\left(\omega-2i\eta\gamma/3\right)^{3}$ and
taking a new origin of time at $t_{B}+4\eta\gamma^{2}/9b^{3}$, we
get: 
\begin{equation}
H_{B}(z,\omega)\approx\exp\left[-\frac{i}{3\eta b^{3}}\left(\omega-\frac{2}{3}i\eta\gamma\right)^{3}-\frac{\eta^{2}}{3}\left(\frac{2\gamma}{3b}\right)^{3}\right].\label{eq:trentedeux}
\end{equation}
 By means of an inverse Fourier transform, we finally find: 
\begin{equation}
h_{B}(z,t)\approx B\:\mathrm{Ai}\left(-\eta^{1/3}bt"\right)\exp\left(-2\eta\gamma t"/3\right).\label{eq:trentetrois}
\end{equation}
 Here $B=\eta^{1/3}b\:\exp\left[-(\eta^{2}/3)\left(2\gamma/3b\right)^{3}\right]$,
$t"=t-t_{B}-4\eta\gamma^{2}/9b^{3}$ and $\mathrm{Ai}(s)$ designates
the Airy function. The range of validity of Eq.(\ref{eq:trentetrois})
can be roughly estimated by means of a strategy similar to that used
for the Sommerfeld precursor. By taking account of the cumulants $k_{4}$
(correction of the attenuation) and $k_{5}$ (correction of the dispersion),
the transfer function associated with the Brillouin precursor approximately
reads as $H_{B}(z,\omega)\times\left(1-a_{4}\omega^{4}-ia_{5}\omega^{5}\right)$
where $a_{4}=-k_{4}/24>0$ and $a_{5}=k_{5}/120>0$ . $H_{B}(z,\omega)$
will be a good approximation if $a_{4}\omega^{4}$ and $a_{5}\omega^{5}$
are small compared to $1$ (say $\leq1/\sqrt{10}$). For sake of simplicity,
we take for the ratios $\omega_{p}/\omega_{0}$ and $\gamma/\omega_{0}$
the values retained by Brillouin, representative of a dense Lorentz
medium with moderate damping. We get then $\eta\approx1.018\approx1$.
Besides, in a cavalier manner, we assimilate $\omega$ to the instantaneous
frequency derived from the asymptotic form $\mathrm{Ai}(-s)\approx\pi^{-1/2}s^{-1/4}\sin\left(2s^{3/2}/3+\pi/4\right)$
that provides a good approximation of $\mathrm{Ai}(-s)$ when $s>1$.
We get so $\omega\approx\sqrt{b^{3}t"}$. With all these hypotheses,
we finally find that the corrections due to the cumulants $k_{4}$
and $k_{5}$ will be small if $\omega_{0}t"\leq2\left(\omega_{0}/b\right)^{3/2}$
and $\omega_{0}t"\leq\left(\omega_{0}/b\right)^{9/5}$, respectively.
Despite the roughness of the procedure leading to these conditions,
it will appear below that they are realistic and even too severe.

\subsection{Precursor generated by the canonical incident field}

When $h_{B}(z,t)$ is slowly varying compared to $e(0,t)$, the Brillouin
originated by the canonical incident field $\sin\left(\omega_{c}t\right)\mathrm{u}_{H}(t)$
takes again the simple form $e_{B}(z,t)=\mathcal{A\:}h_{B}(z,t)$,
that is 
\begin{equation}
e_{B}(z,t)\approx\frac{B}{\omega_{c}}Ai\left(-\eta^{1/3}bt"\right)\exp\left(-2\eta\gamma t"/3\right).\label{trentequatre}
\end{equation}
 It is assumed by writing Eq.(\ref{trentequatre}) that the instantaneous
frequency $\sqrt{b^{3}t"}$ is small compared to $\omega_{c}$ (say
$\sqrt{b^{3}t"}\leq\omega_{c}/\sqrt{10}$) and that the conditions
of validity of $h_{B}(z,t)$ are met. All these restrictions are summarized
by the inequality 
\begin{equation}
\omega_{0}t"\leq\min\left[2\left(\omega_{0}/b\right)^{3/2},\:\left(\omega_{0}/b\right)^{9/5},\:\omega_{0}\omega_{c}^{2}/10b^{3}\right].\label{eq:trentecinq}
\end{equation}
\begin{figure}[h]
\begin{centering}
\includegraphics[width=80mm]{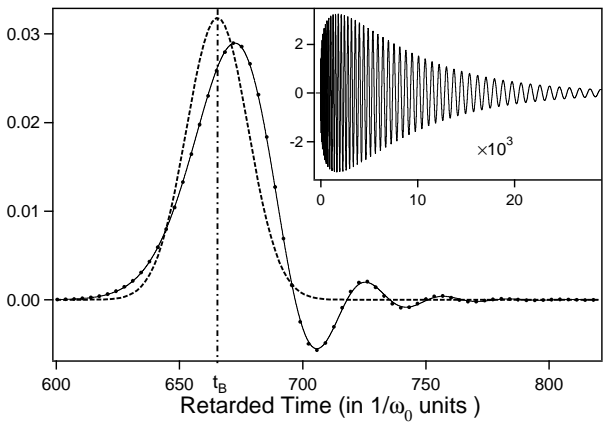} 
\par\end{centering}
\caption{Brillouin precursor obtained in the simple asymptotic limit with the
canonical incident field $\sin(\omega_{c}t)u_{H}(t)$. Parameters
(in $\omega_{0}$ units): $\omega_{c}=1$, $\omega_{p}=1.11$, $\gamma=0.0707$
and $\xi=831$, leading to $\omega_{0}t_{B}\approx665.4$, $b\approx8.44\times10^{-2}\omega_{0}$
and $\beta\approx5.64\times10^{-2}\omega_{0}$. The solid line, the
bullets $\bullet$ and the dashed line are respectively the exact
numerical solution, the analytical solution given Eq.(\ref{trentequatre})
and the Gaussian that would be obtained in the dominant-attenuation
approximation. The conditions are those of Fig.\ref{fig:Sommerfeld_sin}.
The corresponding Sommerfeld precursor is given in inset for reference.\label{fig:BrillouinXi831}}
\end{figure}
Fig.\ref{fig:BrillouinXi831} shows the Brillouin precursor obtained
in the simple asymptotic limit considered in the study of the Sommerfeld
precursor (Fig.\ref{fig:Sommerfeld_sin}). The inequality of Eq.(\ref{eq:trentecinq})
then leads to $\omega_{0}t\leq\min\left[750,\:760,\:840\right]$ .
Insofar as the amplitude of the precursor is negligible for $\omega_{0}t=750$,
the analytical expression of Eq.(\ref{trentequatre}) perfectly fits
the exact numerical result.

\begin{figure}[h]
\begin{centering}
\includegraphics[width=80mm]{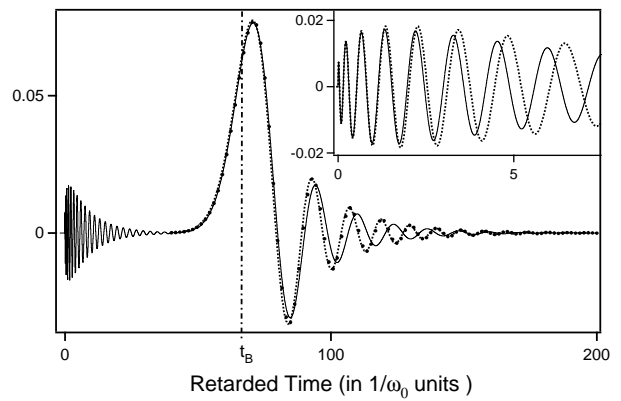} 
\par\end{centering}
\caption{Comparison of the Brillouin precursor obtained outside the asymptotic
limit (solid line) with the analytical forms given Eq.(\ref{trentequatre})
($\bullet$) and Eq.(\ref{eq:trentesix}) (dashed line). Parameters
(in $\omega_{0}$ units): $\omega_{c}=1$, $\omega_{p}=1.11$, $\gamma=0.0707$
and $\xi=83.1$, leading to $\omega_{0}t_{B}\approx66.54$, $b\approx0.182\omega_{0}$
and $\beta\approx0.178\omega_{0}$. Inset: corresponding Sommerfeld
precursor (solid line) compared to the analytic form given Eq.(\ref{eq:vingt})
(dashed line).\label{fig:BrillouinXi83}}
\end{figure}

Surprisingly enough, Eq.(\ref{trentequatre}) remains a not too bad
approximation of the exact result even when the opacity region is
not broad in the sense given to this expression in the present paper.
Fig.\ref{fig:BrillouinXi83} shows the precursor obtained at a distance
ten times smaller than the previous one. Though the width of the opacity
region is then of the order of $\omega_{0}$ {[}see curve (e) of Fig.\ref{fig:GainCurve}{]},
the entirety of the first oscillation of the Brillouin precursor is
very well fitted by Eq.(\ref{trentequatre}). The corresponding Sommerfeld
precursor (inset) is itself well reproduced by Eq.(\ref{eq:vingt})
up to its maximum.

\subsection{Dominant-dispersion limit}

The expression of the Brillouin precursor given by Eq.(\ref{trentequatre})
obviously includes as particular case the Gaussian obtained in the
dominant-attenuation limit. In fact, retrieving the Gaussian precursor
directly from Eq.(\ref{trentequatre}) requires long and tedious calculations
and this probably explains why the Gaussian solution has been generally
overlooked. An other particular form of Eq.(\ref{trentequatre}),
also of special importance, is that obtained when the damping is very
small, so that the formation of the Brillouin precursor is mainly
governed by the group delay dispersion (dominant-dispersion limit).
This requires in particular that $\gamma\ll b$. We then get $t"\approx t-t_{B}$,
$B\approx b$ and 
\begin{equation}
e_{B}(z,t)\approx\frac{b}{\omega_{c}}Ai\left[-b\left(t-t_{B}\right)\right]\exp\left[-\frac{2}{3}\gamma\left(t-t_{B}\right)\right].\label{eq:trentesix}
\end{equation}
 Except for the exponential damping term, this result was established
by Brillouin himself by means of the method of stationary phase \cite{bri32,re4}.
When the group-delay dispersion is fully dominant (say when $\gamma/b<1/100$),
the precursor has a well marked oscillatory behavior with a very weak
damping and its maximum practically coincides with the first maximum
of $Ai\left[-b\left(t-t_{B}\right)\right]$, attained for $t-t_{B}\approx1,02/b$.
The corresponding amplitude is $a_{B}\approx0.536\left(b/\omega_{c}\right)$
that scales as $z^{-1/3}$, instead of as $z^{-1/2}$ in the strict
or dominant-attenuation limit. 
\begin{figure}[h]
\begin{centering}
\includegraphics[width=80mm]{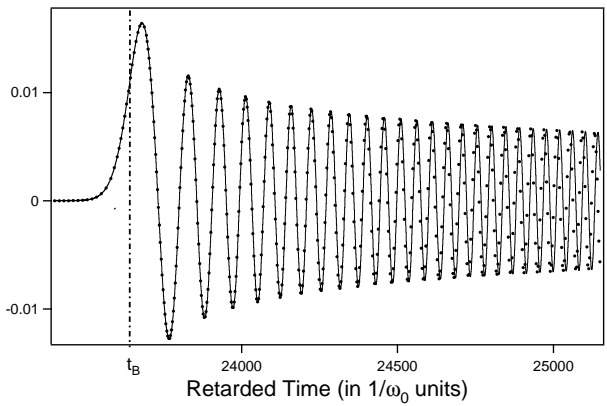} 
\par\end{centering}

\caption{Brillouin precursor in the dispersion dominant limit. The solid line
(bullets $\bullet$) is the exact numerical solution (the analytical
solution). Parameters (in $\omega_{0}$ units): $\omega_{c}=0.836$,
$\omega_{p}=1.11$, $\gamma=10^{-4}$ and $\xi=2.95\times10^{4}$,
leading to $\omega_{0}t_{B}\approx2.3641\times10^{4}$, $b\approx0.0257\omega_{0}$
and $\beta\approx0.252\omega_{0}$. The carrier frequency $\omega_{c}$
is at the lower boundary of the opacity region {[}$\alpha\left(\omega_{c}\right)z\approx20$
{]}.\label{fig:BrillouinDominantDispersion}}
\end{figure}
Fig.\ref{fig:BrillouinDominantDispersion} shows an example of Brillouin
precursor obtained in such conditions ($\gamma/b\approx3.9\times10^{-3}$).
It is worth emphasizing that, since $b\varpropto z^{-1/3}$, the condition
$\gamma/b\ll1$ requires that the propagation distance is not too
large. On the other hand, it should be large enough for the inequality
of Eq.(\ref{eq:trentecinq}) to be satisfied for a time larger or
at least comparable to the half-maximum duration of the precursor.
In fact, the most severe restriction originates in the condition $\omega_{0}\left(t-t_{B}\right)\leq\left(\omega_{0}/b\right)^{9/5}$
associated with the dispersion correction. When $\gamma\ll b$, we
easily deduce from the asymptotic form of the Airy function that the
half-maximum of the precursor will be attained for $\omega_{0}\left(t-t_{B}\right)\approx20\left(\omega_{0}/b\right)$.
The precursor will thus be well reproduced by the expression $e_{B}(z,t)\approx\left(b/\omega_{c}\right)Ai\left[-b\left(t-t_{B}\right)\right]$
beyond its half-maximum amplitude if $\gamma\ll b$ and if $\left(\omega_{0}/b\right)^{4/5}>20$,
that is if $b/\omega_{0}<0.024$. The latter condition is approximately
met Fig.\ref{fig:BrillouinDominantDispersion} for which $b/\omega_{0}=0.026$.
As expected, the maximum amplitude of the precursor is $a_{B}\approx0.536\left(b/\omega_{c}\right)\approx0.0165$
, with $\exp\left(-2\gamma\left(t-t_{B}\right)/3\right)\approx0.997$
at the corresponding time.

\section{PROPAGATION OF PULSES WITH A SQUARE OR GAUSSIAN ENVELOPE\label{sec:SQUARE-GAUSS-PULSES} }

Up to now, in the spirit of the pioneering work of Sommerfeld and
Brillouin, we have considered incident fields of infinite duration.
In actual or even numerical experiments, this duration is naturally
finite. As a matter of fact the simulations made to corroborate our
previous analytical calculations were made by using a square-wave
modulation (eventually suitably filtered) and choosing a square duration
long enough to avoid that the precursors generated by the rise and
the fall of the square overlap. On the contrary, we consider in this
section the case where the duration of the incident field is small
compared to the time-delay $t_{B}$ separating the Brillouin precursor
from the Sommerfeld precursor and does not exceed few periods of the
carrier. We will restrict the analysis to the Brillouin precursor.
Indeed the Sommerfeld precursor, if it exists, is generally much smaller
and will be often filtered out by rise-time effects, to which the
Brillouin precursor is much less sensitive.

\subsection{Square pulse}

\begin{figure}[h]
\begin{centering}
\includegraphics[width=80mm]{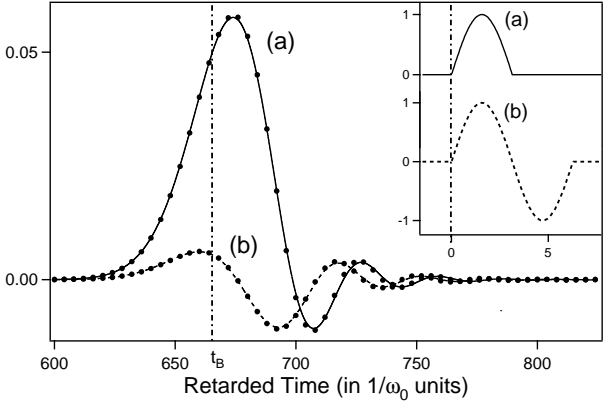} 
\par\end{centering}

\caption{Comparison of the Brillouin precursors $e'_{B}(z,t)$ generated by
an incident square-modulated field of duration (a) $T=T_{c}/2$ and
(b) $T=T_{c}$ . The parameters are those of Fig.\ref{fig:BrillouinXi831}.
The solid and dashed lines are the exact numerical solutions, indiscernible
from the analytical solutions given by Eq.(\ref{eq:trentesept}).
The bullets are the approximate solutions (a) $2e_{B}(z,t-T_{c}/4)$
and (b) $T_{c}\dot{e}_{B}(z,t-T_{c}/2$ . As expected the precursor
amplitude for $T=T_{c}/2$ is twice that attained with a step-modulated
field (see Fig.\ref{fig:BrillouinXi831}) whereas that attained for
$T=T_{c}$ is much smaller. Inset: corresponding incident fields.\label{fig:BrillouinImpulsionCarre}}
\end{figure}

We consider first a square-modulated incident field $\left[\mathrm{u}_{H}(t)-\mathrm{u}_{H}(t-T)\right]\sin\left(\omega_{c}t\right)$.
Of particular interest is the case where the square duration is an
integer $n$ of half-periods of the carrier, that is $T=nT_{c}/2=n\pi/\omega_{c}$.
The incident field can then be rewritten as $e(0,t)=\mathrm{u}_{H}(t)\sin\left(\omega_{c}t\right)-\left(-1\right)^{n}\mathrm{u}_{H}(t-T)\sin\left[\omega_{c}\left(t-T\right)\right]$
and the transmitted field reads as $e'(z,t)=e(z,t)-\left(-1\right)^{n}e(z,t-T)$
where $e(z,t)$ designates the transmitted field when only the incident
field $\mathrm{u}_{H}(t)\sin\left(\omega_{c}t\right)$ is on. This
equation applies to the whole field and in particular to the Brillouin
precursor to yield: 
\begin{equation}
e_{B}'(z,t)=e_{B}(z,t)-\left(-1\right)^{n}e_{B}(z,t-T),\label{eq:trentesept}
\end{equation}
 where $e_{B}(z,t)$ is given by Eq.(\ref{eq:vingtsix}) or Eq.(\ref{trentequatre}),
depending on the system and the parameters considered. The two components
of $e_{B}'$ are of opposite (same) sign when $n$ is even (odd) and
are well separated when it is large enough, so that $T$ significantly
exceeds the duration of the elementary precursor. On the other hand,
$e_{B}(z,t)$ evolving slowly at the scale of $T_{c}$, the two components
overlap and interfere if $n$ is small. When $n=2$ ($T=T_{c}$) as
considered in \cite{ou05,ou90}, the two components interfere nearly
destructively to give a precursor $e_{B}'(z,t)\approx T_{c}\dot{e}_{B}(z,t-T_{c}/2)$.
The case where $n$ is odd and, in particular, where $n=1$ ($T=T_{c}/2$)
is much more favorable. Indeed the two precursors then interfere constructively
to yield a precursor $e_{B}'(z,t)\approx2e_{B}(z,t-T_{c}/4)$ whose
amplitude is twice that obtained with a step modulation. This result
is not really a surprise since the pulse area is itself twice that
of $\mathrm{u}_{H}(t)\sin\left(\omega_{c}t\right)$ . On the contrary
the pulse area equals zero when $n$ is even. The previous results
are illustrated Fig.\ref{fig:BrillouinImpulsionCarre} that shows
the Brillouin precursors obtained for $n=1,2$ for a Lorentz medium
when attenuation and dispersion comparably contribute to the formation
of the Brillouin precursor (simple asymptotic limit).

When the detection of the Brillouin precursor is not time-resolved
an important parameter is the integrated ``energy\textquotedblright{}
$W_{B}(z)=\intop_{-\infty}^{+\infty}\left|e'_{B}(z,t)\right|^{2}dt$
\cite{choi04,lu09}. Thanks to the Parseval-Plancherel theorem \cite{pap87},
it can be written as 
\begin{equation}
W_{B}(z)=\frac{1}{2\pi}\intop_{-\infty}^{+\infty}\left|H_{B}(z,\omega)\right|^{2}\left|E(0,\omega)\right|^{2}d\omega.\label{eq:trentehuit}
\end{equation}
 In this expression all phases are eliminated and $\left|H_{B}(z,\omega)\right|^{2}$
is reduced to $\exp\left(-4\gamma\omega^{2}/3b^{3}\right)=\exp\left(-\omega^{2}/2\beta^{2}\right)$
in both strict and simple asymptotic cases. For $T=T_{c}/2$, $\left|H_{B}(z,\omega)E(0,\omega)\right|^{2}\approx\left(4/\omega_{c}^{2}\right)\exp\left(-\omega^{2}/2\beta^{2}\right)$
and we get an energy $W_{B}(z)=2^{3/2}\pi^{-1/2}\beta/\omega_{c}^{2}$
which slowly decays with the propagation distance (as $1/\sqrt{z}$).
On the other hand, for $T=T_{c}$, $\left|H_{B}(z,\omega)E(0,\omega)\right|^{2}\approx\left(2\pi\omega/\omega_{c}^{2}\right)^{2}\exp\left(-\omega^{2}/2\beta^{2}\right)$
and $W_{B}(z)=(2\pi)^{3/2}\beta^{3}/\omega_{c}^{4}$. As expected,
$W_{B}(z)$ then decays very rapidly with the propagation distance
(as $z^{-3/2}$). As already mentioned, the previous expressions of
the energy are valid regardless of the relative contributions of the
absorption and the dispersion to the formation of the precursor. For
the Debye medium and the Lorentz medium in the dominant-attenuation
limit, it is besides possible to derive from Eq.(\ref{eq:trentesept})
and Eq.(\ref{eq:vingtsix}) explicit expressions of the maximum amplitude
$a_{B}'(z)$ of the precursor $e_{B}'(z,t)$. We find that this amplitude,
equal to $2\beta/\left(\omega_{c}\sqrt{\pi}\right)\approx1.1\left(\beta/\omega_{c}\right)\propto1/\sqrt{z}$
when $T=T_{c}/2$, falls down to $2\sqrt{2\pi/\mathrm{e}}\left(\beta/\omega_{c}\right)^{2}\approx3.0\left(\beta/\omega_{c}\right)^{2}\propto1/z$
when $T=T_{c}$.

\subsection{Gaussian pulse}

The Gaussian pulses are probably the sole smooth pulses for which
it is possible to obtain exact analytic expressions of the Brillouin
precursor, both in the strict and simple asymptotic limit. Non-chirped
incident fields of the form $\mathrm{e}^{-t^{2}/T^{2}}\cos\left(\omega_{c}t\right)$
and $\mathrm{e}^{-t^{2}/T^{2}}\sin\left(\omega_{c}t\right)$ have
been respectively considered by Oughstun and Balictsis in \cite{ou96}
and by Ni and Alfano in \cite{ni06}. When the pulses are linearly
chirped, it is convenient to consider them as the real and imaginary
part of $\widetilde{e}(0,t)=\exp\left(i\omega_{c}t-t^{2}/T^{2}+i\chi^{2}t^{2}\right)$
where $\chi^{2}$ is the chirping parameter. The Fourier transform
of $\widetilde{e}(0,t)$ and of the corresponding transmitted field
$\widetilde{e}_{B}(z,t)$ simply read as $\widetilde{E}(0,\omega)=\widetilde{T}\sqrt{\pi}\exp\left[-\left(\omega-\omega_{c}\right)^{2}\widetilde{T}^{2}/4\right]$
and 
\begin{equation}
\widetilde{E}_{B}(z,\omega)=\widetilde{\mathcal{A}}H_{B}(z,\omega)\exp\left(-\omega^{2}\widetilde{T}^{2}/4+\omega\omega_{c}\widetilde{T}^{2}/2\right).\label{trenteneuf}
\end{equation}
 In these expressions $\widetilde{T}=T/\sqrt{1-i\chi^{2}T^{2}}$ and
$\widetilde{\mathcal{A}}=\widetilde{T}\sqrt{\pi}\exp\left(-\omega_{c}^{2}\widetilde{T}^{2}/4\right)$
may be respectively seen as the (complex) duration and area of the
pulse $\widetilde{e}(0,t)$. In the strict asymptotic limit {[}see
Eq.(\ref{eq:vingtquatre}){]}, we get 
\begin{equation}
\widetilde{E}_{B}(z,\omega)=\widetilde{\mathcal{A}}\exp\left[-\frac{\omega}{4}^{2}\left(\frac{1}{\beta^{2}}+\widetilde{T}^{2}\right)+\omega\frac{\omega_{c}\widetilde{T}^{2}}{2}\right],\label{eq:quarante}
\end{equation}
 and $\widetilde{e}_{B}(z,t)$, inverse Fourier transform of $\widetilde{E}_{B}(z,\omega)$,
reads as 
\begin{multline}
\widetilde{e}_{B}(z,t)=\frac{\widetilde{\mathcal{A}}\beta}{\sqrt{\pi\left(1+\beta^{2}\widetilde{T}^{2}\right)}}\\
\times\exp\left[-\frac{\beta^{2}\left(t'-i\omega_{c}\widetilde{T}^{2}/2\right)}{1+\beta^{2}\widetilde{T}^{2}}\right]\label{eq:quaranteetun}
\end{multline}
 where $t'=t-t_{B}$. In the simple asymptotic limit (see Sec. \ref{sec:EXTENDED-EXPRESSION-BP}),
Eq.(\ref{eq:trenteetun}) and Eq.(\ref{trenteneuf}) yield 
\begin{multline}
\widetilde{E}_{B}(z,\omega)=\widetilde{\mathcal{A}}\exp\left[-i\omega\left(t_{B}+\frac{i\omega_{c}\widetilde{T}^{2}}{2}\right)\right]\\
\times\exp\left[-\omega^{2}\left(\frac{2\gamma}{3b^{3}}+\frac{\widetilde{T}^{2}}{4}\right)-i\omega^{3}\left(\frac{1}{3\eta b^{3}}\right)\right]\label{eq:quarantedeux}
\end{multline}
 This equation is easily transformed in an equation similar to Eq.(\ref{eq:trentedeux}).
By this way, we find 
\begin{equation}
\widetilde{e}_{B}(z,t)=\widetilde{\mathcal{A}}\:\widetilde{B}\:\mathrm{Ai}\left(-\eta^{1/3}b\:\widetilde{t}\right)\exp\left(-\frac{2}{3}\eta\widetilde{\gamma}\:\widetilde{t}\right),\label{eq:quarantetrois}
\end{equation}
 where $\widetilde{\gamma}=\gamma+3b^{3}\widetilde{T}^{2}/8$, $\widetilde{B}=\eta^{1/3}b\:\exp\left[-(\eta^{2}/3)\left(2\widetilde{\gamma}/3b\right)^{3}\right]$
and $\widetilde{t}=t-t_{B}-4\eta\widetilde{\gamma}^{2}/9b^{3}-i\omega_{c}\widetilde{T}^{2}/2$.
Finally the precursors generated by the incident fields $\mathrm{e}^{-t^{2}/T^{2}}\cos\left(\omega_{c}t+\chi^{2}t^{2}\right)$
and $\mathrm{e}^{-t^{2}/T^{2}}\sin\left(\omega_{c}t+\chi^{2}t^{2}\right)$
respectively read as $e_{cos}(z,t)=\Re\left[\widetilde{e}_{B}(z,t)\right]$
and $e_{sin}(z,t)=\Im\left[\widetilde{e}_{B}(z,t)\right]$. Eq.(\ref{eq:quaranteetun}),
Eq.(\ref{eq:quarantetrois}) and the derived expressions of $e_{cos}(z,t)$
and $e_{sin}(z,t)$ hold whatever the duration of the incident pulse
may be. However, as shown below, the amplitude of the Brillouin precursor
will be only significant when this duration does not exceed a few
periods of the carrier. In the Fourier transform $H_{B}(z,\omega)\widetilde{E}(0,\omega)$
of the transmitted field, $H_{B}(z,\omega)$ is then again much narrower
than $\widetilde{E}(0,\omega)$, which may be approximated by its
first order expansion in powers of $\omega$. We get so $\widetilde{E}_{B}(z,\omega)\approx\widetilde{\mathcal{A}}\left(1+\omega\omega_{c}\widetilde{T}^{2}/2\right)H_{B}(z,\omega)$
and finally 
\begin{equation}
\widetilde{e}_{B}(z,t)\approx\widetilde{\mathcal{A}}\left[h_{B}(z,t)-\left(i\omega_{c}\widetilde{T}^{2}/2\right)\dot{h}_{B}(z,t)\right].\label{eq:quarantequatre}
\end{equation}
 When there is no chirping, $\widetilde{T}$ and $\widetilde{\mathcal{A}}$
are real, with $\widetilde{T}=T$ and $\widetilde{\mathcal{A}}=\mathcal{A}=T\sqrt{\pi}\exp\left[-\omega_{c}^{2}T^{2}/4\right]$.
Eq.(\ref{eq:quarantequatre}) then leads to 
\begin{multline}
e_{cos}(z,t)\approx\mathcal{A}\: h_{B}(z,t)\\
=T\sqrt{\pi}\exp\left[-\frac{\omega_{c}^{2}T^{2}}{4}\right]\: h_{B}(z,t)\label{eq:quarantecinq}
\end{multline}
 
\begin{equation}
e_{sin}(z,t)\approx-\frac{\mathcal{A}\omega_{c}T^{2}}{2}\:\dot{h}_{B}(z,t)=-\frac{\omega_{c}T^{2}}{2}\:\dot{e}_{cos}(z,t).\label{eq:quarantesix}
\end{equation}
 
\begin{figure}[h]
\begin{centering}
\includegraphics[width=80mm]{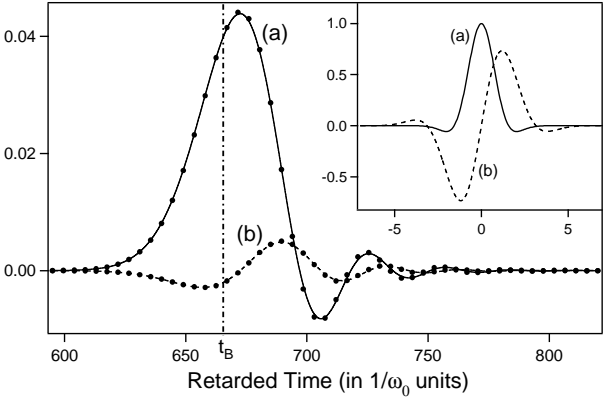}
\par\end{centering}

\caption{Brillouin precursors generated by the incident fields of Gaussian
envelope (a) $\mathrm{e}^{-(t/T)^{2}}\mathrm{cos}(\omega_{c}t)$ with
$T=\sqrt{2}/\omega_{c}$ and (b) $\mathrm{e}^{-(t/T)^{2}}\mathrm{sin}(\omega_{c}t)$
with $T=\sqrt{6}/\omega_{c}$. The parameters are those of Fig.\ref{fig:BrillouinXi831}.
In both cases, the pulse duration has been chosen in order to maximize
the precursor amplitude (see text). The solid and dashed lines are
the exact numerical solutions whereas the bullets are the analytical
solutions obtained in the short pulse approximation {[}Eq.(\ref{eq:quarantecinq})
and Eq.(\ref{eq:quarantesix}){]}, indiscernible from those obtained
without approximation {[}Eq.(\ref{eq:quarantetrois}){]}. Inset: corresponding
incident fields. Numerical calculations shows that the Sommerfeld
precursors generated by these fields have negligible amplitudes, respectively
(a) $5.6\times10^{-7}$ and (b) $1.15\times10^{-10}$. \label{fig:BrillouinImpulsionGaussienne}}
\end{figure}
As illustrated Fig.\ref{fig:BrillouinImpulsionGaussienne}, obtained
in the simple asymptotic limit, these approximate analytic solutions
perfectly fit the exact numerical solution. It is easily deduced from
Eq.(\ref{eq:quarantecinq}) {[}Eq.(\ref{eq:quarantesix}){]} that
the amplitude of the precursor $e_{cos}(z,t)$ {[}$e_{sin}(z,t)${]}
is maximum for a pulse duration $T=T_{m}=\sqrt{2}/\omega_{c}$ {[}
$\sqrt{6}/\omega_{c}${]}. The energy of the precursors can be obtained
by the method already used in the case of a square modulation. We
get so $W_{B}\approx\left(\pi/2\right)^{1/2}\left(\beta T^{2}e^{-\omega_{c}^{2}T^{2}/2}\right)\propto1/\sqrt{z}$
for $e(0,t)=\mathrm{e}^{-t^{2}/T^{2}}\cos\left(\omega_{c}t\right)$
and $W_{B}\approx\left(\pi/32\right)^{1/2}\left(\beta^{3}\omega_{c}^{2}T^{6}e^{-\omega_{c}^{2}T^{2}/2}\right)\propto z^{-3/2}$
for $e(0,t)=\mathrm{e}^{-t^{2}/T^{2}}\sin\left(\omega_{c}t\right)$.
In fact the scaling laws in $z^{-1/2}$ or $z^{-3/2}$ are general
and hold for every short incident pulse. In all cases, the transmitted
pulse is indeed proportional to $h_{B}(z,t)$ when $E(0,0)=\mathcal{A}\neq0$
or to $\dot{h}_{B}(z,t)$ when $\mathcal{A}=0$, the proportionality
coefficient depending only on the characteristics of the incident
pulse and not on the propagation distance. For Gaussian incident pulses
and, more generally, for smooth pulses, the amplitude and the energy
of the Brillouin precursor rapidly decreases with the pulse duration.
For example, the amplitude of the Brillouin precursor generated by
the incident field $\mathrm{e}^{-t^{2}/T^{2}}\cos\left(\omega_{c}t\right)$
is reduced by a factor exceeding $400$ when $T$ is taken four times
larger than its optimum value $\sqrt{2}/\omega_{c}$ {[}see Eq.(\ref{eq:quarantecinq}){]}.
This reduction of amplitude can however be compensated by using chirped
pulses. When the pulse duration remains small enough, Eq.(\ref{eq:quarantequatre})
holds and the Brillouin precursor generated by the incident field
$\mathrm{e}^{-t^{2}/T^{2}}\cos\left(\omega_{c}t+\chi^{2}T^{2}\right)$
reads as 
\begin{multline}
e_{B}(z,t)\approx h_{B}(z,t)\:\Re\left(\widetilde{\mathcal{A}}\right)\\
-\dot{h}_{B}(z,t)\:\Re\left(i\omega_{c}\widetilde{\mathcal{A}}\widetilde{T}^{2}/2\right).\label{eq:quarantesept}
\end{multline}
 Anticipating that the second term of this equation is small compared
to the first one, we easily get the approximate expression 
\begin{equation}
e_{B}(z,t)\approx\mathcal{A}\: h_{B}\left[z,t"-\Re\left(i\omega_{c}\widetilde{\mathcal{A}}\widetilde{T}^{2}/2\mathcal{A}\right)\right],\label{eq:quarantehuit}
\end{equation}
where $\mathcal{A}=\Re\mathcal{\left(\widetilde{A}\right)}$ is the
area of the incident pulse. This result differs from that obtained
without chirping {[}see Eq.(\ref{eq:quarantecinq}){]} by a extra
time-delay $\Re\left(i\omega_{c}\widetilde{\mathcal{A}}\widetilde{T}^{2}/2\mathcal{A}\right)$
and, moreover, by the pulse area $\mathcal{A}$ that may be considerably larger than that attained when the pulse
is not chirped. 
\begin{figure}[h]
\begin{centering}
\includegraphics[width=80mm]{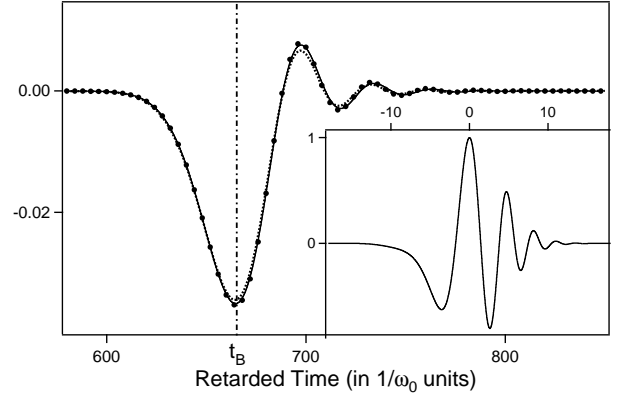} 
\par\end{centering}
\caption{Brillouin precursor generated by a chirped incident pulse $e^{-\left(t/T\right)^{2}}\cos\left(\omega_{c}t+\chi^{2}t^{2}\right)$,
with $T=4\sqrt{2}/\omega_{c}$ and $\chi=\omega_{c}/4$ . The other
parameters are as those of Fig.\ref{fig:BrillouinXi831} and Fig.\ref{fig:BrillouinImpulsionGaussienne}.
The solid line, the bullets and the dashed line respectively are the
exact numerical solution, the analytic solution derived from Eq.(\ref{eq:quarantetrois})
and the approximate analytic solution of Eq.(\ref{eq:quarantehuit}),
obtained in the short pulse approximation. Inset: incident pulse.
The corresponding Sommerfeld precursor has fully negligible amplitude
($9\times10^{-11}$ !).\label{fig:BrillouinChirpedGaussianPulse}}
\end{figure}
Fig.\ref{fig:BrillouinChirpedGaussianPulse} shows the result obtained
for a pulse duration $T=4\sqrt{2}/\omega_{c}$. In order to maximize
the precursor amplitude, we have chosen for the chirping the value
$\chi=\omega_{c}/4$ for which the function $\mathcal{A}(\chi)$ reaches
its first extremum (negative minimum). For these parameters, $\Re\left(i\mathcal{A}\widetilde{T}^{2}/2\mathcal{A}\right)$
is also negative (time advancement). We remark that, despite the numerous
approximations having led to Eq.(\ref{eq:quarantehuit}), it provides
a very good approximation of the exact result.

\section{Conclusion\label{sec:SUMMARY} }

We have analytically studied the propagation of light pulses in a
dense Lorentz medium at distances $z$ so large that the medium is opaque
in a broad spectral region and the Sommerfeld and Brillouin precursors
are far apart from each other. 

Assuming that the carrier frequency $\omega_{c}$ lies in the opacity
region (below, inside or beyond the anomalous dispersion region),
we have shown that the Sommerfeld precursor has a shape independent
of $\omega_{c}$ and that it is entirely determined by the order $p$
and the importance $d_{p}$ of the initial discontinuity of the incident
field, regardless of its subsequent evolution. When the incident field
is discontinuous ($p=0$), its amplitude is independent of $z$ and
$\omega_{c}$ . For $p>0$, this amplitude is proportional to $\omega_{c}z^{-(2p+1)/4}$
and rapidly decreases with the rise time of the incident field. These
results, exact in the strict asymptotic limit where $z\rightarrow\infty$,
provide excellent approximations for the propagation distance considered
by Brillouin and remain good approximations even when $z$ is 1 000
times shorter. 

In the strict asymptotic limit, the formation of the Brillouin precursor
is uniquely determined by the frequency dependence of the medium attenuation.
When $\omega_{c}$ lies in the opacity region, we have shown that
the Brillouin precursor is a Gaussian of amplitude $a_{B}\varpropto1/\left(\omega_{c}\sqrt{z}\right)$
or a Gaussian-derivative of amplitude $a_{B}\varpropto1/\left(\omega_{c}^{2}z\right)$,
depending whether the area of the incident field differs or not from
zero. We have also determined the transmitted field when $\omega_{c}$
is outside the opacity region, evidencing the \textquotedblleft{}pollution\textquotedblright{}
of the Brillouin precursor by the field that is then transmitted at
$\omega_{c}$ (Fig.7).

In a simple asymptotic limit, both attenuation and group delay dispersion
contribute to the formation of the Brillouin precursor. We have established
in this case an expression of the Brillouin precursor containing as
particular cases the previous one (dominant-attenuation limit) and
that obtained by Brillouin by means of the stationary phase method
(dominant-dispersion limit).

We have finally obtained exact analytical expressions of the Brillouin
precursors originated by pulses of square or Gaussian envelope. We
have in particular determined the pulse parameters optimizing the
precursor amplitude and demonstrated that the \emph{energy} of the precursor
decreases with the propagation distance as slowly as $z^{-1/2}$ when
the area of the incident field differs from zero but as rapidly as
$z^{-3/2}$ in the contrary case. We have also shown that, for a given
duration, the precursor amplitude can be greatly enhanced by using
frequency-chirped pulses.

Our explicit analytic expressions of the precursors contrast by their
simplicity from those currently derived by the uniform saddle point
methods. The complexity of the latter \cite{ou09} is often such that
it is difficult and sometimes impossible to retrieve from them our
asymptotic forms. On the other hand, it should be kept in mind that
our results only hold in the limit where the medium is opaque in a
spectral region whose width is much larger than the resonance frequency.
We however remark that they provide a not too bad reproduction of
the Sommerfeld and Brillouin precursors even when this width is of
the order of the resonance frequency (Fig.\ref{fig:BrillouinXi83}).
We finally mention that the study of the precursors is greatly simplified
when the complex index of the medium is such that  $\lvert\widetilde{n}(\omega)-1\rvert\ll1$
$\forall\omega$ \cite{bm11}. As in the study
of the quasi-resonant precursors \cite{bm09}, the equation giving
the saddle points can then be reduced to a biquadratic form and the
saddle point method is expected to provide simple solutions even when
the Sommerfeld and Brillouin precursors partially overlap. This work
is in progress.


\begin{thebibliography}{10}
\bibitem{re2} Keep in mind that we use a retarded time picture where
the time $t$ is the real time minus the luminal propagation time
$z/c$ .
\bibitem{re3} As usual, we consider that the zero-order derivative
of a function is the function itself.
\bibitem{re4} Note that the Airy function $\mathcal{A}(x)$ used
by Brillouin differs from that currently used $\mathrm{Ai}(x)$, with
$2\pi\mathrm{Ai}(-x)=3^{1/3}\mathcal{A}(3^{1/3}x)$. Peak amplitudes
of $\mathcal{A}(x)$ and $\mathrm{Ai}(-x)$ are respectively $2.33$
and $0.536$.
\end{thebibliography}
\end{document}